\documentclass[11pt,a4paper]{article} 
\usepackage{jheppub}
\pdfoutput=1
\usepackage{graphicx}
\usepackage{amsmath}
\usepackage{amssymb}
\usepackage{subcaption}
\usepackage{bbm} 
\usepackage{hhline} 
\usepackage{cleveref} 
\usepackage{tikz-feynman}

\newcommand{\msbar}{\overline{\mathrm{MS}}}

\newcommand{\bare}{\mathrm{bare}}

\newcommand{\cA}{{\mathcal A}}
\newcommand{\cM}{{\mathcal M}}

\newcommand{\cO}{{\mathcal O}}
\newcommand{\cL}{{\mathcal L}}

\newcommand{\cT}{{\mathcal T}}

\newcommand{\cTcut}{\cT_{\mathrm{cut}}}
\newcommand{\cTS}{{\cT_S}}

\newcommand{\cut}{\mathrm{cut}}

\newcommand{\cusp}{\mathrm{cusp}}

\newcommand{\Idiv}{I_{\mathrm{div.}}}
\newcommand{\Ireg}{I_{\mathrm{reg.}}}

\title{Analytical results for the C-angularity soft function at NNLO}

\author[a,b]{Alexander Bennett,}
\author[a]{Emmet P. Byrne,}
\author[a,c]{Jonathan R. Gaunt,}
\author[a,d]{Elsa C. Lang}
\affiliation[a]{Department of Physics and Astronomy, The University of Manchester, Manchester M13 9PL, UK}
\affiliation[b]{Institute for Particle Physics Phenomenology, Durham University, Durham DH1 3LE, UK}
\affiliation[c]{Department of Physics, University of Cyprus, Nicosia 1678, Cyprus}
\affiliation[d]{School of Mathematics, Trinity College, Dublin, Ireland}

\emailAdd{alexander.bennett@durham.ac.uk}
\emailAdd{Emmet.Byrne@manchester.ac.uk}
\emailAdd{gaunt.jonathan@ucy.ac.cy}
\emailAdd{elang@tcd.ie}

\abstract{
We compute the soft function at NLO and NNLO for a one-parameter family of event shapes we call C-angularity. This family contains  C-parameter as a specific choice of the parameter, in close analogy with how conventional angularity contains thrust as a special case. By construction, C-angularity and angularity coincide in the collinear limit such that the anomalous dimensions are equal. However, unlike angularity, C-angularity is a continuously differentiable function of the final state momenta, which makes the analytic calculation of the C-angularity soft function simpler. We obtain analytical results for the C-angularity soft function and anomalous dimension as an expansion in the C-angularity parameter $a$, to third and fourth order in $a$ respectively. These expansions yield results that are accurate at the few per mille level for $-1\le a < 1$.
}


\begin{document}

\begin{flushright}
  IPPP/25/74
\end{flushright}
\vspace{1cm}

\maketitle

\section{Introduction}

Event shapes are simple observables with a long history. They are functions of the momenta of all the hadrons (or partons) in the final state, and therefore event shapes characterise the geometry of the hadronic final state without requiring a definition of a jet algorithm. Classic examples include thrust~\cite{Brandt:1964sa,Farhi:1977sg}, C-parameter~\cite{Fox:1978vu,Fox:1978vw,Ellis:1980wv} and jet-broadening~\cite{Rakow:1981qn,Catani:1992jc,Dokshitzer:1998kz}. Event shapes have been
measured to high accuracy at electron-positron colliders such as LEP and PETRA, and have been used for precise determinations of the strong coupling constant, $\alpha_s$~\cite{Hoang:2015hka,Abbate:2010xh,Abbate:2012jh}. They are also useful tools to discriminate between the gluonic and fermionic decay modes of the Higgs, at future leptonic Higgs factories \cite{Gao:2016jcm, Mo:2017gzp, Gao:2019mlt, Luo:2019nig, Ju:2023dfa, Knobbe:2023njd, Wang:2023azz, Zhu:2023oka, Yan:2023xsd, Gehrmann-DeRidder:2024avt, Fox:2025qmp, Fox:2025txz, Fox:2025vxj}.

Two-jet event shapes, $\tau_e$, are distributions which are dominated by two almost back-to-back collimated sets of particles in the limit $\tau_e\to 0$. A broad class of event shapes in $e^+ e^- \to X$ is given by 
\begin{equation}
    \tau_e(X) = \frac{1}{Q}\sum_{i \in X} f_e(\eta_i)|\vec{p}_{i\perp}|\,,
    \label{eq:taue}
\end{equation}
where $Q$ is the centre-of-mass energy, and $\eta_i$ and $\vec{p}_{i\perp}$ are the rapidity and transverse momentum of particle $i$ in the hadronic final state $X$, with respect to the thrust axis of the event\footnote{Other axes may be considered, for example the broadening axis~\cite{Larkoski:2014uqa} or the winner-take-all axis~\cite{Bertolini:2013iqa}.}. Thrust, broadening, and C-parameter\footnote{This differs to the standard Lorentz-invariant event shape of ref.~\cite{Ellis:1980wv,Gardi:2003iv} but the definitions coincide in the dijet limit, up to a factor of 6~\cite{Salam:2001bd}.} are given by a choice of
\begin{equation}
    f_\tau(\eta)=e^{-|\eta|}\,, \qquad 
    f_b(\eta)=1\,, \qquad 
    f_C(\eta)=\frac{1}{2 \cosh \eta}\,,
\end{equation}
respectively. We note that thrust and C-parameter coincide in the forward and backward limits, $\eta \to \pm \infty$, which, in particular, ensure the anomalous dimensions for these observables coincide. 

In the dijet region $\tau_e \ll 1$, the cross section differential in $\tau_e$ contains large logarithms $\ln(\tau_e)$ which must be resummed to all orders to achieve a stable, reliable prediction. A high order of resummation accuracy is required for precise extractions of $\alpha_s$; state-of-the-art predictions for thrust are at N$^3$LL$^\prime$ order \cite{Abbate:2012jh, Benitez:2024nav}, with results at a similar level of accuracy for C-parameter \cite{Hoang:2014wka}.

Angularity~\cite{Berger:2003iw,Berger:2003pk}, $\tau_{\tau,a}$, is a family of event shapes with
\begin{equation}
    f_{\tau,a}(\eta)=e^{-|\eta|(1-a)}\,, 
    \label{eq:fta}
\end{equation}
which is an infrared-safe observable for $-\infty < a < 2$, and which reduces to thrust for $a=0$ and to jet-broadening in the limit $a\to1$. Angularity has been computed to NNLL$^\prime$ accuracy utilising a numerical calculation of the next-to-next-to-leading order (NNLO) soft function~\cite{Bell:2018vaa,Bell:2018oqa,Bell:2018gce}. More recently, the NNLO soft anomalous dimension for angularity has also been computed analytically as an expansion in $a$ up to $\mathcal{O}(a^2)$~\cite{Bauer:2020npd}. 

\begin{figure}
    \centering
    \includegraphics[width=0.7\linewidth]{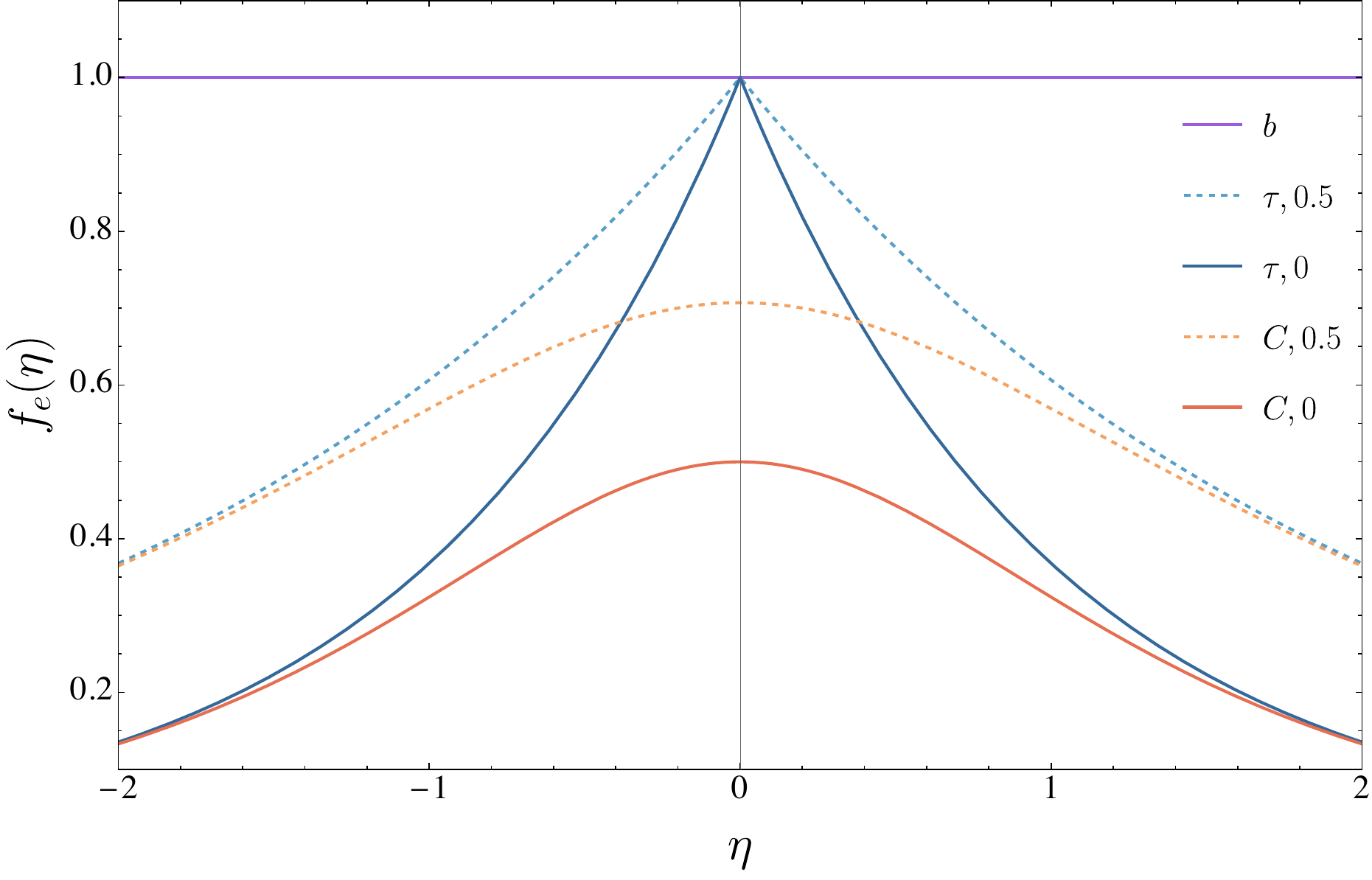}
    \caption{A comparison of the function, $f_e(\eta)$ for several of the dijet event shapes in the class defined by \cref{eq:taue}.  In particular, we compare $f_e$ for broadening ($e=b$), angularity ($e=\tau,a$) and C-angularity ($e=C,a$). For angularity, we plot $f_{\tau,a}$ for $a=0$ (i.e. thrust) and $a=0.5$. For C-angularity, we plot $f_{C,a}$ for $a=0$ (i.e. C-parameter) and $a=0.5$.}
    \label{fig:eventshapes}
\end{figure}

In this work, we focus on an alternative family of event shapes, which corresponds to a generalisation of C-parameter rather than thrust. We call this \emph{C-angularity}, $\tau_{C,a}$, and define this by choosing
\begin{equation}
    f_{C,a}(\eta)=\left(\frac{1}{2 \cosh \eta}\right)^{1-a}\,.
    \label{eq:fCa}
\end{equation}
This event shape has not been discussed in the literature before, apart from a brief mention in ref.~\cite{Lee:2006nr}. Just as for angularity, \cref{eq:fCa} defines an infrared-safe observable for $-\infty < a < 2$. At $a=0$, C-angularity reduces to C-parameter, while in the limit $a\to1$ it tends to jet-broadening. As is evident from \cref{eq:fta,eq:fCa}, angularity and C-angularity coincide in the forward and backward limits (that is, $\eta \to \pm \infty$), which ensures the anomalous dimensions for these observables coincide for any $a$. This feature is apparent in \cref{fig:eventshapes}, in which we have plotted $f_e$ for several event shapes. While the discussion here and in the following will be couched in the context of $e^+ e^-$ event shapes, C-angularity may also be studied in the context of deep inelastic scattering (DIS), analogous to the recent study of angularity in DIS performed in ref.~\cite{Zhu:2021xjn}.

The study of such (C-)angularity event shapes is of interest in the context of extractions of $\alpha_S$. For a precise extraction, one needs to understand the impact on the event shape from nonperturbative (or hadronisation) effects. If one is in the tail region of the distribution (but still in a region dominated by dijet configurations, $\tau_e \simeq 0$), and neglecting the effect of final state hadron masses, the impact is determined by a universal nonperturbative parameter $\Omega_1$ scaled by exactly calculable observable-dependent coefficients\footnote{For the case of our C-angularities, this coefficient $c_{C,a}$ is given by $B((1-a)/2,1/2)/2^{(1-a)}$, where $B(x,y)$ is the Beta function, whilst for angularities the coefficient $c_{\tau,a} = 2/(1-a)$ \cite{Berger:2003pk, Berger:2004xf, Lee:2006nr}. \label{footnote_pccoeff}} \cite{Webber:1994cp, Dokshitzer:1995qm, Dokshitzer:1995zt, Dokshitzer:1998pt, Korchemsky:1998ev,  Korchemsky:1999kt, Gardi:2001ny, Gardi:2002bg, Lee:2006fn, Lee:2006nr, Becher:2013iya}. Then, extracting $\alpha_S$ generally involves a 2D fit of both $\alpha_S$ and $\Omega_1$, and the use of data/predictions at different $a$ values can be useful to break degeneracies in this fitting procedure \cite{Bell:2018gce}. In recent years there has been a considerable amount of research activity and discussion on the topic of nonperturbative power corrections to event shapes \cite{Salam:2001bd, Mateu:2012nk, Luisoni:2020efy, Caola:2021kzt, Caola:2022vea, Bell:2023dqs, Nason:2023asn, Banfi:2023mes, Benitez:2024nav, Dasgupta:2024znl, Nason:2025qbx, Farren-Colloty:2025amh, Banfi:2025crj} -- C-angularities could offer an additional avenue to study the power corrections and test these predictions. Note that C-angularities are affected by power corrections in a different way to the conventional angularities (as can be seen from footnote \ref{footnote_pccoeff}) -- this is because the dominant power corrections are driven by emissions in the soft wide-angle region, and C-angularity has a different shape to angularity here.

As C-angularity is a new class of event shapes, there is no existing experimental data to which we can currently compare our results. Due to the distinct rapidity dependence of \cref{eq:fCa}, the measurement of C-angularity would complement the measurement of other well-established event shapes at future $e^+ e^-$ colliders, such as the FCC-ee~\cite{FCC:2018evy}, CEPC~\cite{CEPCStudyGroup:2018ghi}, CLIC~\cite{CLICdp:2018cto} or ILC~\cite{ILCInternationalDevelopmentTeam:2022izu}. More immediately, there is an ongoing collaborative effort to re-analyse and make available archived data from past $e^+ e^-$ colliders~\cite{Bossi:2025nux}. This allows for the possibility to perform a measurement of C-angularity from archived data, in a similar manner to the recent measurements of thrust using archived data from DELPHI~\cite{Zhang:2025nlf} and ALEPH~\cite{Badea:2025wzd,Electron-PositronAlliance:2025hze} experiments.

In this paper we derive the next-to-leading order (NLO) and NNLO soft function for C-angularity, which are required for the resummation of C-angularity up to the NNLL$^\prime$/N$^3$LL order. We emphasise the computational simplicity of the derivation of our results. The calculation of the thrust or angularity soft function requires the partition of the real-emission phase-space into two hemispheres, which complicates the theoretical calculation. For example, sophisticated techniques were required to extend the reverse unitarity method to handle piecewise-defined observables such as thrust or zero-jettiness~\cite{Baranowski:2021gxe}. Likewise, the numerical calculation of thrust and angularity soft functions have been reported to be prone to numerical instabilities, which C-parameter does not suffer from~\cite{Bell:2018oqa}. 

The calculation of the NNLO C-angularity soft function presented here can be checked by taking several different limits. In the limit $a\to0$ we recover the NNLO C-parameter soft function first computed in ref.~\cite{Hoang:2014wka} using a fit to results from the \texttt{EVENT2} generator~\cite{Catani:1996jh,Catani:1996vz}, and more recently computed analytically in ref.~\cite{Bell:2018oqa}. The limit $a\to2$ reproduces the known NNLO threshold soft function \cite{Belitsky:1998tc,Becher:2006nr}, which is relevant for high-mass colour singlet production in proton-proton collisions \cite{Sterman:1986aj, Catani:1989ne, Ravindran:2006bu, Westmark:2017uig, Banerjee:2017cfc, Banerjee:2018vvb}. Using our results, we also confirm the (C-)angularity anomalous dimension computed in ref.~\cite{Bauer:2020npd} (up to a numerically tiny disagreement in the analytic results for the $a^2$ term of the $C_RC_A$ colour channel), and we extend these results by two powers in $a$.

The main principle underlying the present analytic calculation of the C-angularity soft function at NNLO is a separation into a soft function for a simpler (but infrared-unsafe) observable which we term \emph{global C-angularity}, plus a correction term. The calculation of the NNLO global C-angularity soft function can be straightforwardly extracted from a kinematic limit of a previously computed beam function. The correction term has a much simpler divergence structure which enables the coefficients of an expansion in $a$ to be performed analytically to a high order in $a$. A similar approach to computing soft and beam functions has been used in refs.~\cite{Bauer:2020npd,Jouttenus:2011wh,Banfi:2014sua,Gangal:2016kuo,Abreu:2022sdc, Abreu:2022zgo, Abreu:2026FutureJV, Buonocore:2026FutureTau}.

The method of calculation presented in this work can be readily extended to compute the C-angularity soft function at N$^3$LO, where such results would also include the N$^3$LO result for the C-parameter soft function (taking $a \to 0$) and the N$^3$LO (C-)angularity anomalous dimension. The anomalous dimension would be the last missing piece to resum (C-)angularities to N$^3$LL, whilst the result for the soft function is required for the resummation of C-angularities at N$^3$LL$^\prime$/N$^4$LL (and in the case $a=0$, it is the last missing piece required for full N$^3$LL$^\prime$ accuracy).

The structure of the rest of the paper is as follows. In section \ref{sec:bg} we discuss the factorisation of the C-angularity event shape, and define the soft function appearing in this factorisation formula. In section \ref{sec:NLOcalc} we present the NLO calculation of this soft function. Section \ref{sec:NNLOcalc} describes the strategy we employ for the NNLO calculation of this quantity, and presents the NNLO results. Finally, in section \ref{sec:conclusion}, we conclude.

\section{C-angularity}
\label{sec:bg}

The factorisation of differential distributions of event shapes of the form \cref{eq:taue} is proven in refs.~\cite{Bauer:2008dt, Hornig:2009vb} using Soft Collinear Effective Theory (SCET)~\cite{Bauer:2000ew,Bauer:2001yt,Bauer:2000yr,Bauer:2001ct}. As C-angularity belongs to this class of event shapes, the factorisation of the distribution for $\tau_{C,a} \ll 1$ has the form
\begin{align}
\begin{split}
    \frac{1}{\sigma^{(0)}}\frac{\mathrm{d}\sigma}{\mathrm{d} \tau_{C,a}}= \frac{H(Q,\mu)}{Q^2}
    \int& 
    \mathrm{d} \cT_S \,
    \mathrm{d} \cT_1 \,
    \mathrm{d} \cT_2 \,
    \delta(Q \,\tau_{C,a}-\cT_1-\cT_2-\cT_S)\\
    &
    \times S_a(\cT_S,\mu)
    \,
    J_a(Q, \cT_1,\mu)
    \,
    J_a(Q, \cT_2,\mu)\,.
    \label{eq:fac}
\end{split}
\end{align}
We have opted to write the C-angularity soft function, $S_a$, and the jet functions $J_a$ in terms of variables $\cT_i$ with mass-dimension 1. 
Note that this formula receives corrections associated with recoil effects; as $a \to 0$ (and for negative $a$) these are suppressed by powers of $\tau_{C,a}$, but as $a \to 1$ these corrections become of the same order as the right hand side of eq.~\eqref{eq:fac} itself (see ref.~\cite{Budhraja:2022fqz}).
Due to the fact that angularity and C-angularity coincide in the collinear limit, $J_a$ are identical to the angularity jet functions computed numerically at one loop in ref.~\cite{Hornig:2009vb} and at two loops in refs.~\cite{Bell:2018gce,Bell:2021dpb}. The corresponding factorisation formula for C-angularities in DIS has one jet function replaced by an angularity beam function $B_a$, which is computed at one loop in ref.~\cite{Zhu:2021xjn} and at two loops in ref.~\cite{Bell:2024lwy}.

The (bare version of the) soft function in \cref{eq:fac} is defined in the usual way as a matrix element of two light-like Wilson lines, with the measurement $\delta(\cT_{C,a} - \cT_S)$ applied to the final-state soft particles produced (see e.g.~refs.~\cite{Bauer:2008dt, Hornig:2009vb}). This is the differential soft function -- we can also define a cumulant soft function by integrating this:
\begin{equation}
 S^\bare_a(\cTcut)  = \int_0^{\cTcut} \mathrm{d}  \cTS \, S^\bare_a(\cTS)\,.
 \label{eq:cumulant}
\end{equation}
In order to avoid cluttered notation, in the following we will only distinguish the differential and cumulant soft functions by use of the respective argument $\cTS$ and $\cTcut$ for $S_a$.
%

The differential soft function is renormalised via a convolution,
\begin{equation}
S_a^\bare(\cTS)= \left[Z_a^S\otimes S_a\right](\cTS,\mu)\,.
\label{eq:ren_dif}
\end{equation}
where we use the shorthand notation $\otimes$ to denote the convolution 
\begin{equation}
\left[f \otimes g \right](\cTS,\mu)= \int_0^{\cTS} \mathrm{d} \cT' f(\cTS-\cT',\mu)g(\cT',\mu)\,.
\end{equation}
We expand all quantities in $\alpha_s/(4\pi)$, as detailed in \cref{sec:expansion}. At leading order, we have
\begin{equation}
S_a^{(0)}(\cTS,\mu)= \delta(\cTS)\,,
\qquad 
Z_a^{S(0)}(\cTS,\mu)=\delta(\cTS)\,.
\end{equation}
The soft anomalous dimension is determined via 
\begin{equation}
\mu \frac{\mathrm{d}}{\mathrm{d} \mu}S_a(\cTS,\mu)= \left[\gamma_{a;S}\otimes S_a\right](\cTS, \mu)\,,
\label{eq:gamma_a_S_dif}
\end{equation}
or equivalently, via
\begin{equation}
\left[Z^S_a\otimes 
\gamma_{a;S}\right](\cTS, \mu)=
-\mu \frac{\mathrm{d}}{\mathrm{d} \mu}Z^S_a(\cTS,\mu)\,.
\label{eq:gamma_a_S_Z}
\end{equation}
We decompose the soft anomalous dimension into a cusp and non-logarithmic piece, as in ref.~\cite{Bauer:2020npd},
\begin{equation}
\gamma_{a;S}(\cTS, \mu)=\frac{4}{1-a}
\Gamma_\cusp\left[\alpha_s(\mu)\right]
\frac{1}{\mu}\cL_0\bigg(\frac{\cTS}{\mu}\bigg)
+
\hat{\gamma}_{a;S}[\alpha_s(\mu)]\delta(\cTS)\,.
\label{eq:gamma_a_S_dif_split}
\end{equation}
Although the soft anomalous dimension depends on the representation of the Wilson lines in the matrix element, we suppress this label for ease of notation.
The quantity $\Gamma_\cusp$ is the cusp anomalous dimension, which is now known up to four loops \cite{Henn:2019swt, Henn:2019rmi, Moch:2018wjh, Moch:2017uml, Davies:2016jie}. In our calculation, we only need (and we reproduce) $\Gamma_\cusp$ up to two loops, the results for which are given in Appendix \ref{sec:expansion}.

In the practical calculation of the C-angularity soft function, it is useful to express the measurement in terms of lightcone coordinates, defined relative to the thrust axis (or the directions of the Wilson lines in the soft function). For an individual massless parton $i$ in the final state, with momentum $k_i$, it is useful to define the function
\begin{align}
\cT_{C,a}(k_i)=\frac{(k_i^+ k_i^-)^{1-\frac{a}{2}}}{(k_i^+ + k_i^-)^{1-a}}\,,
\label{eq:singleparticleC}
\end{align}
We define the lightcone components using the conventions given in appendix \ref{sec:kin}. Then the dimension-$1$ C-angularity appearing in the soft function of \cref{eq:fac} corresponds to
\begin{equation}
\cT_{C,a}=\sum_i  \cT_{C,a}(k_i)\,,
\label{eq:TCP}
\end{equation}
For the special cases of $a=0,1$ and $2$, the single particle contribution to the C-angularity, \cref{eq:singleparticleC}, become:
\begin{subequations}
\begin{align}
    \cT_{C,0}(k_i)&=\frac{k_i^+ k_i^-}{k_i^+ + k_i^-}\,,\\
    \cT_{C,1}(k_i)&=(k_i^+ k_i^-)^\frac{1}{2}=k_{iT}\,,\\
    \cT_{C,2}(k_i)&=k_i^+ + k_i^- \, =2 k_i^0\,. \label{eq:thresh}
\end{align}
\end{subequations}
From this we see that the C-angularity becomes C-parameter, broadening and threshold for $a=0,1$ and $2$ respectively.

Having introduced the necessary properties of the C-angularity soft function, we can proceed to calculate it. 

\section{Calculation at NLO}
\label{sec:NLOcalc}

The NLO bare differential soft function in $\msbar$ is given by
\begin{align}
   S_a^{\bare(1)}(\cTS)&= 4 C_R \, (4 \pi)^2 \, 
   \left(\frac{\mu^2 e^{\gamma_E}}{4\pi}\right)^\epsilon
   \int \frac{\mathrm{d}^{d}k}{(2\pi)^d}
   \frac{(2\pi)\delta(k^2)\theta(k^0)}{k^+ k^-}\delta(\cT_{C,a}-\cTS)
   \label{eq:S1_bare_int}\\
   &
   =
   \frac{8 C_R e^{\epsilon \gamma_E}}{\Gamma(1-\epsilon) }  \int \frac{\mathrm{d}k^+ \mathrm{d}k^-}{2k^+ k^-}
   \left(\frac{\mu^2}{k^+ k^-}\right)^{\epsilon}
   \delta\left(\frac{(k^+ k^-)^{1-\frac{a}{2}}}{(k^++k^-)^{1-a}} -\cTS\right)\,,
   \label{eq:S1_bare_int_2}
\end{align}
where we have integrated over all but the light-cone momenta in the second equality. Note that we have not specified a representation for the final-state Wilson lines. For Wilson lines in the fundamental or adjoint representation we take $C_R$ to be $C_F$ or $C_A$ respectively.
To perform the remaining integrals in \cref{eq:S1_bare_int_2}, a natural change of variables is
\begin{equation}
    k^0=\frac{1}{2}(k^+ + k^-), \qquad t=k^+k^-,
    \label{eq:cov_q0t}
\end{equation}
where $k^0$ is the energy of real radiation and $t$ is sometimes called the transverse virtuality. 
Inverting these relations, we have
\begin{align}
\max (k^+,k^-) = k^0 + \sqrt{(k^0)^2-t}, \qquad
\min (k^+,k^-)  = k^0 - \sqrt{(k^0)^2-t},
\end{align}
which makes it clear that $(k^0)^2\ge t$ for $k^+,k^- \in \mathbb{R}$. Exploiting the symmetry of the integrand in $k^+ \leftrightarrow k^-$, we can impose $k^+>k^-$ then double the result. Doing so, we obtain
\begin{align}
S_a^{\bare(1)}(\cTS)&=
   \frac{8 C_R e^{\epsilon \gamma_E}}{\Gamma(1-\epsilon) }  \int \frac{\mathrm{d}k^0 \mathrm{d}t\, (2k^0)^{1-a}}{t \sqrt{(k^0)^2-t}}\theta\left((k^0)^2-t\right)
   \left(\frac{\mu^2}{t}\right)^{\epsilon}
   \delta
   \left(\frac{t^{1-\frac{a}{2}}}{(2k^0)^{1-a}} -\cTS\right)\,.
\end{align}
Imposing the $\delta$-function constraint by integrating over $t$, then integrating over $k^0$, we find
\begin{align}
S_a^{\bare(1)}(\cTS)&=
   \frac{4 C_R e^{\epsilon \gamma_E}}{\Gamma(1-\epsilon) } 
   \frac{\Gamma(\epsilon(1-a))^2}{\Gamma(2\epsilon(1-a))}
   \left[
   \frac{1}{\mu}
   \left(\frac{\cTS}{\mu}\right)^{-1-2\epsilon}
   \right]\,,
\end{align}
to all orders in $\epsilon$. Expanding the quantity in square brackets in distributions (see \cref{eq:dist}), we find
\begin{align}
\begin{split}
S_a^{\bare(1)}(\cTS)=
   \frac{4 \, C_R}{1-a} \Bigg(
   &-\frac{\delta(\cTS)}{\epsilon^2}
   +\frac{2}{\epsilon}\frac{1}{\mu}\cL_0\left(\frac{\cTS}{\mu}\right)
   \\
   &-4\frac{1}{\mu}\cL_1\left(\frac{\cTS}{\mu}\right)
   +\frac{\pi^2}{12}(3-4a+2a^2)\delta(\cTS)
   \Bigg)+\cO(\epsilon)\,.
\label{eq:S1_bare}
\end{split}
\end{align}
Note that the Eikonal amplitude in \cref{eq:S1_bare_int} is flat in rapidity. At the value $a=1$ the measurement function also becomes independent of rapidity and the integral generates an unregulated rapidity divergence. This manifests in \cref{eq:S1_bare} as a pole at $a=1$. An additional rapidity regulator such as the CMU~\cite{Chiu:2011qc,Chiu:2012ir} or exponential~\cite{Li:2016axz} regulator would then be required but we do not implement this in our current work. As such, our results are valid for $a\neq 1$, but we recall that recoil corrections are also large (i.e.~of the same order as the singular cross section, \cref{eq:fac}) as $a \to 1$.

Expanding the renormalisation condition \cref{eq:ren_dif} to NLO we find
\begin{align}
Z_a^{S(1)}(\cTS, \mu)&=
\frac{4 \, C_R}{1-a} \left[
   -\frac{\delta(\cTS)}{\epsilon^2}
   +\frac{2}{\epsilon}\frac{1}{\mu}\cL_0\bigg(\frac{\cTS}{\mu}\bigg)
   \right]
   \,, \\
S_a^{(1)}(\cTS, \mu)&=
   \frac{4 \, C_R}{1-a} \left[
   -4\frac{1}{\mu}\cL_1\bigg(\frac{\cTS}{\mu}\bigg)
   +\frac{\pi^2}{12}(3-4a+2a^2)\delta(\cTS)
   \right]+\cO(\epsilon)\,.
\end{align}
After performing a Laplace transform,
\begin{equation}
    \tilde{S}_a(\tau)=\int_0^\infty \frac{\mathrm{d} \cTS}{\mu} \exp\left(-\frac{\cTS}{\mu}e^{-\gamma_E}\tau\right)
    S_a(\cTcut)\,,
    \label{eq:Laplace}
\end{equation}
we find agreement with the existing literature on the C-parameter in the limit $a \to 0$~\cite{Gangal:2014qda,Hoang:2014wka}, and we find agreement with the existing literature on the threshold soft function in the limit $a\to 2$~\cite{Belitsky:1998tc,Idilbi:2006dg,Becher:2006nr,Becher:2007ty}. Expanding \cref{eq:gamma_a_S_dif} to leading order, we find
\begin{align}
\Gamma_{\cusp}^{(0)}&=4 \, C_R\,,\qquad
\hat{\gamma}_{a;S}^{(0)}= 0\,,
\end{align}
in agreement with \cref{eq:Gamma0} and ref.~\cite{Bauer:2020npd}.

For completeness, we list the renormalised  cumulative soft function which, according to \cref{eq:cumulant}, is given by 
\begin{align}
S_a^{(1)}(\cTcut,\mu)&=
   \frac{4 \, C_R}{1-a} 
   \left[
   -2\log\bigg(\frac{\cTcut}{\mu}\bigg)^2
   +\frac{\pi^2}{12}(3-4a+2a^2)
   \right]+\cO(\epsilon)\,.
\end{align}

\section{Calculation at NNLO}
\label{sec:NNLOcalc}

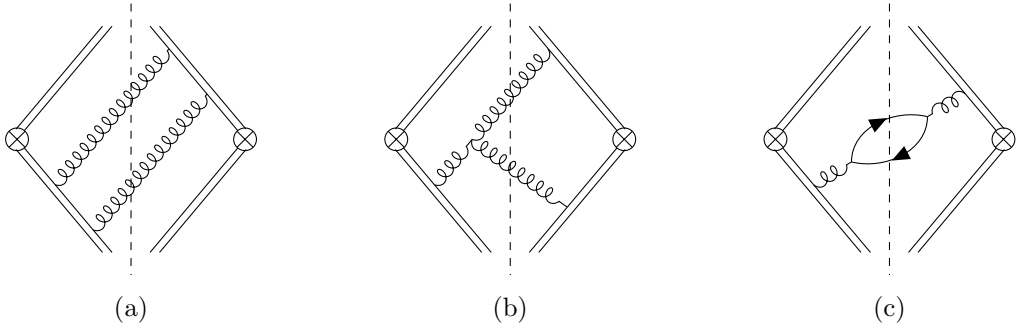
\begin{figure}
\begin{subfigure}{.33\textwidth}
    \centering
    \begin{tikzpicture}
    \begin{feynman}
\def\xscale{0.25} 
\def\yscale{0.3} 
\vertex (i0) at (7*\xscale, 12*\yscale);
\vertex (i1) at (5.5*\xscale, 11*\yscale);
\vertex (i2) at (8*\xscale, 11*\yscale);
\vertex (i3) at (6*\xscale, 11*\yscale);
\vertex (i4) at (8.5*\xscale, 11*\yscale);
\vertex (i5) at (9*\xscale, 10*\yscale);
\vertex (i6) at (11*\xscale, 8*\yscale);
\vertex (i7) [crossed dot] at (1*\xscale, 6*\yscale) {};
\vertex (i8) [crossed dot] at (13*\xscale, 6*\yscale) {};
\vertex (i9) at (0.5*\xscale, 6*\yscale) {};
\vertex (i10) at (13.5*\xscale, 6*\yscale) {};
\vertex (i11) at (3*\xscale, 4*\yscale);
\vertex (i13) at (5*\xscale, 2*\yscale);
\vertex (i14) at (8.5*\xscale, 1*\yscale);
\vertex (i15) at (8*\xscale, 1*\yscale);
\vertex (i16) at (6*\xscale, 1*\yscale);
\vertex (i17) at (5.5*\xscale, 1*\yscale);
\vertex (i18) at (7*\xscale, 0*\yscale);
\diagram* {
(i0) -- [scalar] (i18),
(i1) -- [plain] (i9) -- [plain] (i17),
(i2) -- [plain] (i5) -- [plain] (i6) -- [plain] (i8) -- [plain] (i15),
(i3) -- [plain] (i7) -- [plain] (i11) -- [plain] (i13) -- [plain] (i16),
(i14) -- [plain] (i10) -- [plain] (i4),
(i11) -- [gluon] (i5),
(i13) -- [gluon] (i6)
};
    \end{feynman}
    \end{tikzpicture}
    \caption{}
\end{subfigure}%
\begin{subfigure}{.33\textwidth}
    \centering
    \begin{tikzpicture}
    \begin{feynman}
\def\xscale{0.25} 
\def\yscale{0.3} 
\vertex (i0) at (7*\xscale, 12*\yscale);
\vertex (i1) at (5.5*\xscale, 11*\yscale);
\vertex (i2) at (8*\xscale, 11*\yscale);
\vertex (i3) at (6*\xscale, 11*\yscale);
\vertex (i4) at (8.5*\xscale, 11*\yscale);
\vertex (i5) at (9*\xscale, 10*\yscale);
\vertex (i6) at (0.5*\xscale, 6*\yscale) {};
\vertex (i7) [crossed dot] at (13*\xscale, 6*\yscale) {};
\vertex (i8) [crossed dot] at (1*\xscale, 6*\yscale) {};
\vertex (i9) at (5*\xscale, 6*\yscale);
\vertex (i10) at (13.5*\xscale, 6*\yscale) {};
\vertex (i11) at (3*\xscale, 4*\yscale);
\vertex (i12) at (10*\xscale, 3*\yscale);
\vertex (i13) at (8.5*\xscale, 1*\yscale);
\vertex (i14) at (8*\xscale, 1*\yscale);
\vertex (i15) at (6*\xscale, 1*\yscale);
\vertex (i16) at (5.5*\xscale, 1*\yscale);
\vertex (i17) at (7*\xscale, 0*\yscale);
\diagram* {
(i0) -- [scalar] (i17),
(i1) --  (i6) -- (i16),
(i2) -- [plain] (i5) -- [plain] (i7) -- [plain] (i12) -- [plain] (i14),
(i3) -- [plain] (i8) -- [plain] (i11) -- [plain] (i15),
(i13) -- [plain] (i10) -- [plain] (i4),
(i9) -- [gluon] (i5),
(i9) -- [gluon] (i12),
(i11) -- [gluon] (i9)
};
    \end{feynman}
    \end{tikzpicture}
    \caption{}
\end{subfigure}%
\begin{subfigure}{.33\textwidth}
    \centering
    \begin{tikzpicture}
    \begin{feynman}
\def\xscale{0.25} 
\def\yscale{0.3} 
\vertex (i0) at (7*\xscale, 12*\yscale);
\vertex (i1) at (5.5*\xscale, 11*\yscale);
\vertex (i2) at (8*\xscale, 11*\yscale);
\vertex (i3) at (6*\xscale, 11*\yscale);
\vertex (i4) at (8.5*\xscale, 11*\yscale);
\vertex (i6) at (11*\xscale, 8*\yscale);
\vertex (i7) at (9*\xscale, 7*\yscale);
\vertex (i8) at (13.5*\xscale, 6*\yscale) {};
\vertex (i9) [crossed dot] at (1*\xscale, 6*\yscale) {};
\vertex (i10) [crossed dot] at (13*\xscale, 6*\yscale) {};
\vertex (i11) at (0.5*\xscale, 6*\yscale) {};
\vertex (i12) at (5*\xscale, 5*\yscale);
\vertex (i14) at (3*\xscale, 4*\yscale);
\vertex (i16) at (5.5*\xscale, 1*\yscale);
\vertex (i17) at (6*\xscale, 1*\yscale);
\vertex (i18) at (8*\xscale, 1*\yscale);
\vertex (i19) at (8.5*\xscale, 1*\yscale);
\vertex (i20) at (7*\xscale, 0*\yscale);
\diagram* {
(i0) -- [scalar] (i20),
(i1) -- [plain] (i11) -- [plain] (i16),
(i2) -- [plain] (i6) -- [plain] (i10) -- [plain] (i18),
(i3) -- [plain] (i9) -- [plain] (i14) -- [plain] (i17),
(i19) -- [plain] (i8) -- [plain] (i4),
(i7) -- [fermion, out = -90, in = -14] (i12) -- [fermion, out = 90, in = 166] (i7),
(i6) -- [gluon] (i7),
(i14) -- [gluon] (i12)
};
    \end{feynman}
    \end{tikzpicture}
    \caption{}
\end{subfigure}
\caption{Example two loop soft factor diagrams containing the colour factors (a) $C_R^2$, (b) $C_R C_A$ and (c) $C_R n_f T_F$.}
\label{fig:twoloopdiagrams}
\end{figure}

At NNLO we adopt a similar organisational scheme to ref.~\cite{Gangal:2016kuo}. Primarily, we decompose the soft function into separate colour structures,
\begin{equation}
   S_a^{(2)}=C_R^2 \, S_a^{(2,C_R)}+ C_R C_A  \, S_a^{(2,C_A)} + C_R n_f T_F  \, S_a^{(2, n_f T_F)}\,.
   \label{eq:colour_decomp}
\end{equation}
Example diagrams corresponding to each of these colour factors are given in figure \ref{fig:twoloopdiagrams}.
The `uncorrelated' $C_R^2$ term can be obtained directly from the NLO bare soft function via non-abelian exponentiation \cite{Gatheral:1983cz,Frenkel:1984pz}. This will be performed explicitly in \cref{sec:CRCR}. In \cref{sec:correlated}, we compute the remaining `correlated' colour structures. 
We do this in two steps. In \cref{sec:global} we compute part of the correlated soft function by utilising a kinematic limit of the known NNLO virtuality-dependent beam function~\cite{Gaunt:2014cfa,Gaunt:2014xga}, while in \cref{sec:Delta} we compute the remainder directly from the soft amplitudes given in ref.~\cite{Hornig:2011iu}.

\subsection{\texorpdfstring{$C_R^2$}{CR CR} piece}
\label{sec:CRCR}
The bare $C_R^2$ term can be obtained directly from the self-convolution of the NLO soft function~\cite{Gatheral:1983cz,Frenkel:1984pz}
\begin{equation}
   S_a^{\bare(2,C_R)}=\frac{1}{2}\left[S_a^{\bare(1)} \otimes S_a^{\bare(1)}\right]\,.
   \label{eq:NAE_bare}
\end{equation}
In order to convert \cref{eq:NAE_bare} into an analagous expression for the renormalised beam function, we perform a similar colour decomposition to \cref{eq:colour_decomp} for the renormalisation factor, specifically 
\begin{equation}
   Z_a^{(2)}=C_R^2 \, Z_a^{(2,C_R)}+C_R C_A \, Z_a^{(2,C_A)}+ C_R n_f T_F\, Z_a^{(2,  n_f T_F)}\,.
   \label{eq:colour_decomp_Z}
\end{equation}
We then expand \cref{eq:ren_dif} to the first two orders in $\alpha_s$, whereby we find
\begin{align}
S_a^{(2,C_R)}=\frac{1}{2}\left[S_a^{(1)} \otimes S_a^{(1)}\right]+\frac{1}{2}\left[Z_a^{(1)} \otimes Z_a^{(1)}\right]-Z_a^{(2,C_R)}\,.
\end{align}
This implies
\begin{align}
S_a^{(2,C_R)}=\frac{1}{2}\left[S_a^{(1)} \otimes S_a^{(1)}\right]\,, \qquad
Z_a^{(2,C_R)}=\frac{1}{2}\left[Z_a^{(1)} \otimes Z_a^{(1)}\right]\,.
\end{align}
Utilising the convolutions listed in refs.~\cite{Ligeti:2008ac,Gaunt:2015pea} we find the $C_R^2$ part of the renormalisation function is given by
\begin{align}
Z_a^{(2,C_R)}(\cTS,\mu)=&
\left(\frac{4}{1-a}\right)^2 
\Bigg[
\frac{1}{2\epsilon^4}\delta(\cTS)
-\frac{2}{\epsilon^3}\frac{1}{\mu}\cL_0\bigg(\frac{\cTS}{\mu}\bigg)
+
\frac{4}{\epsilon^2}
\left(\frac{1}{\mu}\cL_1\bigg(\frac{\cTS}{\mu}\bigg)
-\frac{\pi^2}{12}\delta(\cTS)
\right)
\Bigg]\,,
\end{align}
while the $C_R^2$ part of the renormalised soft function is given by
\begin{align}
\begin{split}
S_a^{(2,C_R)}(\cTS,\mu)=&\left(\frac{4}{1-a}\right)^2 
\Bigg[
8\frac{1}{\mu}\cL_3\bigg(\frac{\cTS}{\mu}\bigg)
-\frac{\pi^2}{3}(11-4 a +2 a^2)
\frac{1}{\mu}\cL_1\bigg(\frac{\cTS}{\mu}\bigg)\\
&
\hspace{-4 em}
+16 \zeta_3
\frac{1}{\mu}\cL_0\bigg(\frac{\cTS}{\mu}\bigg)
+\frac{\pi^4}{1440}(13-120 a +140 a^2 - 80 a^3 + 20 a^4)
\delta(\cTS)
\Bigg]+\cO(\epsilon)\,.
\label{eq:Sa2CRCR}
\end{split}
\end{align}
Expanding either \cref{eq:gamma_a_S_dif} or \cref{eq:gamma_a_S_Z} to NNLO we find (as expected)
\begin{align}
\Gamma_{\cusp}^{(1,C_R)}&=0\,,\qquad
\hat{\gamma}_{a;S}^{(1,C_R)}= 0\,,
\end{align}
where we have performed the colour decomposition of \cref{eq:colour_decomp,eq:colour_decomp_Z} also to the anomalous dimension.
Integrating \cref{eq:Sa2CRCR} according to \cref{eq:cumulant} gives us the cumulant soft function,
\begin{align}
\begin{split}
S_a^{(2,C_R)}(\cTcut,\mu)=&\left(\frac{4}{1-a}\right)^2 
\Bigg[
2\log^4\left(\frac{\cTcut}{\mu}\right)
-\frac{\pi^2}{6}(11-4 a +2 a^2)
\log^2\left(\frac{\cTcut}{\mu}\right)\\
&+16 \zeta_3
\log\left(\frac{\cTcut}{\mu}\right)
+\frac{\pi^4}{1440}(13-120 a +140 a^2 - 80 a^3 + 20 a^4)
\Bigg]\,.
\end{split}
\end{align}

\subsection{\texorpdfstring{$C_R C_A$}{CR CA} and \texorpdfstring{$C_R n_f T_F$}{CR nf TF} pieces}
\label{sec:correlated}
We now turn our attention to the correlated colour structures. For these colour structures we simplify the calculation by splitting the measurement function into two pieces. We define the `global' C-angularity as the C-angularity measurement applied to the total momentum of the radiation:
\begin{equation}
    \cT_{C,a}^\Sigma= \cT_{C,a}\left(\textstyle\sum_i k_i\right)\,.
    \label{eq:TCaG}
\end{equation}
In terms of \cref{eq:TCaG}, the original measurement can be written
\begin{equation}
    \cM(\cTcut)=\cM^{\Sigma}(\cTcut)+ \Delta \cM(\cTcut)
    \label{eq:M_split}
\end{equation}
where the correction to the global measurement is
\begin{equation}
    \Delta\cM(\cTcut)=\theta(\cT_{C,a}<\cTcut)
    -\theta(\cT^\Sigma_{C,a}<\cTcut)\,.
    \label{eq:DeltaM}
\end{equation}
The decomposition of the measurement, \cref{eq:M_split}, translates to a decomposition of the soft function, which we denote
\begin{align}
S_a^{\bare(2,c)}(\cTcut) = 
S_{\Sigma,a}^{\bare(2,c)}(\cTcut) + \Delta S_a^{\bare(2,c)}(\cTcut)\,,
\label{eq:S_split}
\end{align}
where $ c \in \{C_A, n_f T_F \} $ are the correlated colour structures.

For a single emission, the global C-angularity measurement clearly coincides with the full measurement, such that real-virtual diagrams do not contribute to $\Delta S_a^{\bare(2,c)}$, and we only have to consider the real-real (RR) ones. Further, for the RR diagrams, the correction to the measurement function, $\Delta \cM_{RR}$, vanishes in most of the soft and collinear regions the of phase space, meaning $\Delta S_a^{\bare(2,c)}$ has a simpler divergence structure than $S_a^{\bare(2,c)}$. The exceptional cases are when both partons become collinear to the same jet direction, e.g.~$k_1^+\sim k_2^+>>k_1^-\sim k_2^-$,
where the full and global C-angularities tend to distinct limits,
\begin{subequations}
\begin{align}
\cT_{C,a}&\to 
(k_1^+)^{\frac{a}{2}}(k_1^-)^{1-\frac{a}{2}}
+
(k_2^+)^{\frac{a}{2}}(k_2^-)^{1-\frac{a}{2}}\label{eq:TPforward}\,,
\\
\cT_{C,a}^\Sigma&\to (k_1^++k_2^+)^{\frac{a}{2}}(k_1^-+k_2^-)^{1-\frac{a}{2}}\label{eq:TGforward}\,,
\end{align}
\end{subequations}
or where one parton is collinear to one jet, and the other parton is collinear to the other, e.g.~$k_1^+\sim k_2^->>k_1^-\sim k_2^+$,
where the full and global C-angularities tend to
\begin{subequations}
\begin{align}
\cT_{C,a}&\to (k_1^+)^{\frac{a}{2}}(k_1^-)^{1-\frac{a}{2}}+(k_2^-)^{\frac{a}{2}}(k_2^+)^{1-\frac{a}{2}}\,,
\label{eq:TPbtb}
\\
\cT_{C,a}^\Sigma&\to (k_1^+)^{1-\frac{a}{2}} (k_2^-)^{1-\frac{a}{2}} (k_1^++k^-_2)^{-1+a}\,.
\label{eq:TGbtb}
\end{align}
\end{subequations}
The latter limit only corresponds to a divergence for the uncorrelated channel -- for the correlated channels discussed here, the amplitude is strongly suppressed in this limit. We note that \cref{eq:TPforward} and \cref{eq:TGforward} do coincide for $a=0$, while \cref{eq:TPbtb} and \cref{eq:TGbtb} remain distinct. To conclude, $\Delta S_a^{\bare(2,c)}$ for $c \in \{ C_A, n_f T_F\}$ has only a single divergence, corresponding to the whole emitted system going to infinite rapidity, and this divergence vanishes as $a\to0$. The decomposition \cref{eq:S_split} is similar to that employed in the calculation of the double-real contribution to the NNLO angularity soft function in ref.~\cite{Bauer:2020npd}, and similar methodology has been employed in refs.~\cite{Jouttenus:2011wh,Banfi:2014sua,Gangal:2016kuo, Abreu:2022sdc, Abreu:2022zgo, Abreu:2026FutureJV, Buonocore:2026FutureTau}. It will be convenient to extend the decomposition \cref{eq:S_split} to the renormalisation function and anomalous dimension, specifically,
\begin{align}
Z_a^{(2,c)}(\cTcut) &= 
Z_{\Sigma,a}^{(2,c)}(\cTcut) + \Delta Z_a^{(2,c)}(\cTcut)\,,\\
\gamma_{a;S}^{(1,c)}(\cTcut) &= 
\gamma_{\Sigma,a;S}^{(1,c)}(\cTcut) + \Delta \gamma_{a;S}^{(1,c)}(\cTcut)\,.
\end{align}
Having established this notation, we can proceed to the calculation of $S_{\Sigma,a}^{(2,c)}$.

\subsubsection{Global C-angularity}
\label{sec:global}
The global C-angularity soft function in \cref{eq:S_split} can be obtained by first integrating over the final-state radiation with a measurement only on the total plus and minus momentum of the radiation:
$$\delta\Big(r^--\sum_i k_i^-\Big)\delta\Big(r^+-\sum_i k_i^+\Big)\,.$$
The remaining integral over $r^+$ and $r^-$ takes the same form as the NLO calculation \cref{eq:S1_bare}, with the replacement $\epsilon \to 2\epsilon$\,,
\begin{align}
\begin{split}
   S_{\Sigma,a}^{\bare(2)}(\cTS)+\frac{\beta_0}{\epsilon} S_{a}^{\bare(1)}(\cTS)=& f(N_c,n_f,\epsilon)
   \\
&\times
\int \frac{\mathrm{d}r^+ \mathrm{d}r^-}{r^+ r^-}
   \left(\frac{\mu^2}{r^+ r^-}\right)^{2\epsilon}
   \delta\left(\frac{(r^+ r^-)^{1-\frac{a}{2}}}{(r^++r^-)^{1-a}} -\cTcut\right)
\end{split}
\\
   =&
   f(N_c,n_f,\epsilon)
   \frac{\Gamma(2\epsilon(1-a))^2}{\Gamma(4\epsilon(1-a))}
   \left[
   \frac{1}{\mu}
   \left(\frac{\cTcut}{\mu}\right)^{-1-4\epsilon}
   \right]\,,
   \label{eq:S2G_bare}
\end{align}
where the structure of $f(N_c,n_f,\epsilon)$ depends on the soft amplitudes which have been integrated over. The simple expression on the right hand side of \cref{eq:S2G_bare} in terms of $f$ only holds when we expand the bare soft function in terms of the bare coupling constant, $\alpha_s^\bare/(4\pi)$, while we expand the bare soft function in the renormalised coupling $\alpha_s/(4\pi)$. Therefore, we must `undo' the coupling renormalisation on the LHS of \cref{eq:S2G_bare}.

We can determine the function $f$ from the endpoint limit $x\to 1$ of the NNLO virtuality-dependent beam function computed in ref.~\cite{Gaunt:2014cfa,Gaunt:2014xga,Boughezal:2017tdd, Baranowski:2020xlp}. In the computation of the beam function, two constraints are placed on the real radiation, $k_i$,
$$\delta\Big((1-x)p^--\sum_i k_i^-\Big)\delta\Big(t-x p^-\sum_i k_i^+\Big)\,,$$
where $p$ is the initial-state momentum and $t=-q^+ q^-$ is the transverse virtuality of the parton entering the hard collision, $q=p-\sum_i k_i$. As discussed in ref.~\cite{Gaunt:2014cfa,Gaunt:2014xga}, the endpoint  limit, $x\to 1$, of the beam function amplitudes can be obtained by replacing the incoming parton lines with collinear Wilson lines, which gives nothing other than the soft amplitudes we would need to compute the soft function (see also refs.~\cite{Billis:2019vxg, Lustermans:2019cau}). By comparison with \cref{eq:S2G_bare}, and identifying
\begin{equation}
    r^+ = (1-x)p^-\,,\qquad r^- = \frac{t}{ xp^-}\,,
\end{equation}
the beam function must have the limit
\begin{align}
&B^{\bare(2)}(t,x)+\frac{\beta_0}{\epsilon}B^{\bare(1)}(t,x)\xrightarrow[x\to 1]{}
f(N_c,n_f,\epsilon)\frac{1}{\mu^2}
\left((1-x)\frac{t}{\mu^2} \right)^{-1-2\epsilon}\,,\\
\begin{split}
&
\qquad 
=
f(N_c,n_f,\epsilon)
\left(
\frac{1}{(-2\epsilon)} \delta(1-x)+\sum_{n=0}^{\infty} \frac{(-2\epsilon)^n}{n!} \cL_n(1-x)
\right)
\\
&
\hspace{7.4 em}
\times
\left(
\frac{1}{(-2\epsilon)} \delta(t)+\sum_{n=0}^{\infty} \frac{(-2\epsilon)^n}{n!}\frac{1}{\mu^2} \cL_n\left(\frac{t}{\mu^2}\right)
\right)\,.
\end{split}
\end{align}
Thus, the function $f$ may be extracted from, e.g.~\cite{Baranowski:2020xlp}. The renormalised soft function is then obtained by expanding \cref{eq:ren_dif} to NNLO, which, for $c \in \{n_f T_F, C_A \}$ simply reads
\begin{equation}
    S_{\Sigma,a}^{\bare(2,c)}(\cTS)=
    Z_{\Sigma,a}^{(2,c)}(\cTS)+S_{\Sigma,a}^{(2,c)}(\cTS)\,.
\end{equation}
Explicitly, the renormalisation factors are
\begin{align}
\begin{split}
Z_{\Sigma,a}^{\bare(2,n_f T_F)}(\cTS,\mu)=\frac{1}{(1-a)}\Bigg[&
\frac{-4}{\epsilon^3}\delta(\cTS)
+\frac{1}{\epsilon^2}
\bigg(
\frac{16}{3}
\frac{1}{\mu}\cL_0\bigg(\frac{\cTS}{\mu}\bigg)
+
\frac{20}{9}
\delta(\cTS)
\bigg)\\
&+\frac{1}{\epsilon}
\bigg(
-\frac{80}{9}
\frac{1}{\mu}\cL_0\bigg(\frac{\cTS}{\mu}\bigg)
+\left(\frac{112}{27}-\frac{2\pi^2}{9}\right)
\bigg)
\Bigg]\,,
\end{split}
\end{align}

\begin{align}
\begin{split}
Z_{\Sigma,a}^{\bare(2,C_A)}(\cTS,\mu)=&\frac{1}{(1-a)}\Bigg[
\frac{11}{\epsilon^3}\delta(\cTS)
+\frac{1}{\epsilon^2}
\bigg(
-\frac{44}{3}
\frac{1}{\mu}\cL_0\bigg(\frac{\cTS}{\mu}\bigg)
+
\left(
-\frac{67}{9}+\frac{\pi^2}{3}
\right)
\delta(\cTS)
\bigg)\\
&+\frac{1}{\epsilon}
\bigg(
\left(
\frac{268}{9}-\frac{4\pi^2}{3}\right)
\frac{1}{\mu}\cL_0\bigg(\frac{\cTS}{\mu}\bigg)
+\left(-\frac{404}{27}+\frac{11\pi^2}{18}+14\zeta_3\right)
\bigg)
\Bigg]\,,
\end{split}
\end{align}
and the renormalised soft functions are
\begin{align}
\begin{split}
S_{\Sigma,a}^{(2,n_f T_F)}(\cTS,\mu)=&\frac{4}{3(1-a)}\Bigg[
-16
\frac{1}{\mu}\cL_2\bigg(\frac{\cTS}{\mu}\bigg)
+\frac{80}{3}
\frac{1}{\mu}\cL_1\bigg(\frac{\cTS}{\mu}\bigg)
\\
&+\frac{4}{3}
\left(-\frac{28}{3}+(2-2a+a^2)\pi^2\right)
\frac{1}{\mu}\cL_0\bigg(\frac{\cTS}{\mu}\bigg)
\\
&+
\left(\frac{164}{27}
-\frac{5}{18}
(7-8a+4a^2)\pi^2
+
2\left(
    \frac{7}{3}-4a(3-3a+a^2)
\right)\zeta_3
\right)\delta(\cTS)
\Bigg]\,,
\label{eq:S_global_nfTF}
\end{split}
\\
\begin{split}
S_{\Sigma,a}^{(2,C_A)}(\cTS,\mu)=&\frac{4}{3(1-a)}\Bigg[
44
\frac{1}{\mu}\cL_2\bigg(\frac{\cTS}{\mu}\bigg)
+4\left(-\frac{67}{3}+\pi^2\right)
\frac{1}{\mu}\cL_1\bigg(\frac{\cTS}{\mu}\bigg)
\\
&+
\left(\frac{404}{9}-\frac{11}{3}(2-2a+a^2)\pi^2-42 \zeta_3\right)
\frac{1}{\mu}\cL_0\bigg(\frac{\cTS}{\mu}\bigg)
\\
&
\hspace{-9 em}
+
\left(-\frac{607}{27}
+\frac{67}{72}
(7-8a+4a^2)\pi^2
-
(1-4a+2a^2)\frac{\pi^4}{12}
-
11\left(
    \frac{7}{6}-2a(3-3a+a^2)
\right)\zeta_3
\right)\delta(\cTS)
\Bigg]\,.\label{eq:S_global_CA}
\end{split}
\end{align}
Expanding \cref{eq:gamma_a_S_dif} or \cref{eq:gamma_a_S_Z} to NNLO, we find
\begin{align}
    \gamma_{\Sigma,a;S}^{(1,n_f T_F)}(\cTS,\mu)&=\frac{4}{(1-a)}\left( 
    -\frac{80}{9}
    \frac{1}{\mu}\cL_0\bigg(\frac{\cTS}{\mu}\bigg)
    +\frac{2}{27}(56-3 \pi^2)\delta(\cTS)
    \right)\,,
    \label{eq:gamma_global_nfTF}\\
    \gamma_{\Sigma,a;S}^{(1,C_A)}(\cTS,\mu)&=\frac{4}{(1-a)}\left( 
    \frac{4}{9}\left(67-3 \pi^2 \right)
    \frac{1}{\mu}\cL_0\bigg(\frac{\cTS}{\mu}\bigg)
    +\frac{1}{54}(-808+33 \pi^2+756\zeta_3)\delta(\cTS)
    \right)\,.
    \label{eq:gamma_global_CA}
\end{align}
We note that the cusp component of the global (C-)angularity soft anomalous dimensions, as defined in \cref{eq:gamma_a_S_dif_split}, reproduces the full two-loop cusp anomalous dimension, \cref{eq:Gamma1}. We therefore expect the correction to the full (C-)angularity soft anomalous dimension will be proportional to $\delta(\cTS)$.

For threshold, $a=2$, global C-angularity,~\cref{eq:TCaG}, coincides with the full C-angularity, \cref{eq:TCP}. Therefore, $\Delta S_a^{(2)}$ vanishes in this limit, and the NNLO threshold soft function should be given by
$$
C_R
\left(
C_F S_{2}^{(2,C_F)}+C_A S_{\Sigma,2}^{(2,C_A)}+n_f T_F S_{\Sigma,2}^{(2,n_f T_F)}\right)\,.$$
We indeed find agreement with the NNLO threshold soft function given in refs.~\cite{Belitsky:1998tc,Becher:2007ty} upon taking the Laplace transform, \cref{eq:Laplace}, of our results.

\subsubsection{Correction from global to full C-angularity}
\label{sec:Delta}
The correction to go from the global to the full C-angularity measurement is given by
\begin{align}
\begin{split}
\Delta S_a^{\bare(2,c)}(\cTcut)=\left(\mu^{2} \frac{e^{\gamma_E}}{4\pi}\right)^{2\epsilon}
   \int 
   \prod_{i \in \{1,2 \}}
   \left[
   \frac{\mathrm{d}^d k_i}{(2\pi)^d}
    (2\pi)\delta(k_i^2)\theta(k_i^0)
    \right]
    \Delta \cM_{RR}(\cTcut) \cA_{RR}^{(2,c)}\,,
\end{split}
\end{align}
where $\cA_{RR}^{(2,c)}$ are the amplitudes for the emission of two soft partons, squared, and summed over final state helicities and colours, which are given in ref.~\cite{Hornig:2011iu}. The superscript indicates this function is the coefficient of $\alpha_s(\mu)^2/(4 \pi)^2$ and colour factor $ c \in \{ C_A, n_f T_F\} $.
Integrating over the transverse degrees of freedom we arrive at
\begin{align}
\begin{split}
    \Delta S_a^{\bare(2,c)}(\cTcut)=
    &
    \frac{e^{2 \epsilon \gamma_E}
    }
    {2^{2 \epsilon}\pi \Gamma(1-2\epsilon)}
    \int_0^\infty
    \prod_{i \in \{1,2 \}}
   \left[
   \frac{\mu^{2\epsilon}
   \mathrm{d}k_i^+\mathrm{d}k_i^- 
   }{
   (k_i^+ k_i^-)^{\epsilon}
   }
    \right]
    \int_0^{\pi} \frac{
    \mathrm{d}(\Delta \phi) 
    }
    {
    \sin^{2\epsilon}(\Delta \phi)
    }
    \Delta \cM_{RR} \cA_{RR}^{(2,c)}\,,
\end{split}
\end{align}
where $\Delta \phi$ is the azimuthal angle between momenta $k_1$ and $k_2$. Next we follow ref.~\cite{Gangal:2016kuo} and change variables to 
\begin{align}
\begin{split}
    y_t &= \frac{1}{4}\log\left(
    \frac{k_1^- k_2^-}{k_1^+ k_2^+}
    \right)\,,\qquad
    \Delta y = \frac{1}{2}
    \log\left(
    \frac{k_1^- k_2^+}{k_1^+ k_2^-}
    \right)\,,
    \\
    T&=k_1^+ + k_2^+\,, \qquad \qquad \qquad
    z=\frac{k_1^+}{k_1^+ + k_2^+}\,.
\end{split}
\end{align}
Integrating over $\Delta \phi$ and $T$ (which on dimensional grounds must be universal across both colour channels) we obtain
\begin{align}
\begin{split}
    \Delta S_a^{\bare(2,c)}(\cTcut)&=
    \left(\frac{\mu}{\cTcut} \right)^{4\epsilon}
    \int_{-\infty}^{\infty}
    \mathrm{d}y_t \, e^{-4 \epsilon y_t}
    \int_{-\infty}^{\infty}
    \mathrm{d}\Delta y
    \int_0^{1}
    \mathrm{d}z
    \,
    \cM(y_t,\Delta y,z)
    \cA^c(\Delta y, z)\,.
\end{split}
\end{align}
Here we have captured the residual dependence from the measurement in the function
\begin{equation}
    \cM(y_t,\Delta y,z)
   =
   \left[
    \frac{(\tau_a^\Sigma)^{4 \epsilon}-(\tau_a)^{4 \epsilon}}{4 \epsilon}
    \right]\,,\\
\end{equation}
with $\tau_a=\cT_{C,a}/T$ and $\tau_a^\Sigma=\cT^\Sigma_{C,a}/T$ (note that in these ratios the dependence on $T$ cancels, leaving a function only of $y_t, \Delta y$ and $z$). We have factored out the $y_t$ dependence from the amplitudes that is common to both colour channels, and we have absorbed all other Jacobian and phase-space factors into the function $\cA^c$, to which we will henceforth refer to as the `amplitude'.

From the discussion in the introduction to \cref{sec:correlated}, we expect the integral over $y_t$ to generate a divergence. To see how this divergence arises, we take the limit of the integrand in the $y_t\to \pm\infty$ limits.
In the forward ($+\infty$) or backward ($-\infty$) limit we find a common $y_t$-dependent function factorises from both global and full terms in the measurement, specifically,
\begin{equation}
\cM(y_t,\Delta y,z) \xrightarrow[y_t \to \pm\infty]{} e^{(1\pm(a-1))y_t} \cM^{\pm}(\Delta y,z)
\end{equation}
where the remaining dependence on $\Delta y$ and $z$ is given by 
\begin{equation}
\cM^{\pm}(\Delta y,z)
   =
   \left[
    \frac{(\tau_a^{\Sigma\pm})^{4 \epsilon}-(\tau_a^{\pm})^{4 \epsilon}}{4 \epsilon}
    \right]\,,
\label{eq:cMpm}
\end{equation}
with constituents
\begin{align}
\tau_a^{+}(\Delta y, z)&=e^{\frac{a}{2}\Delta y}z+e^{-\frac{a}{2}\Delta y}(1-z)\,,
\\
\tau_a^{\Sigma+}(\Delta y, z)&=e^{-\frac{a}{2}\Delta y}
\left(1+(e^{2\Delta y}-1)z
\right)^{\frac{a}{2}}\,,
\\
\tau_a^{-}(\Delta y, z)&=e^{\frac{2-a}{2}\Delta y}z+e^{\frac{a-2}{2}\Delta y}(1-z)\,,
\\
\tau_a^{\Sigma-}(\Delta y, z)&=e^{\frac{a-2}{2}\Delta y}
\left(1+(e^{2\Delta y }-1)z\right)^{\frac{2-a}{2}}\,.
\end{align}
Integrating the forward asymptotic $y_t$ behaviour over positive $y_t$, and the backward asymptotic $y_t$ behaviour over negative $y_t$ leads to equal divergent terms
\begin{align}
\int_0^\infty \mathrm{d} y_t \, e^{-4 y_t (1-a)\epsilon}&=\int_{-\infty}^0 \mathrm{d} y_t \, e^{4 y_t (1-a)\epsilon}=\frac{1}{4(1-a)\epsilon}\,.
\label{eq:DeltaSpoles}
\end{align}
The coefficient of this divergent factor is given by the integral
\begin{align}
I^c_{\mathrm{div.}}=
\int_{-\infty}^{\infty}
    \!
    \mathrm{d}\Delta y
    \int_0^{1}
    \!
    \mathrm{d}z
    \,
    \cM^{\pm}(\Delta y,z)
    \cA^c(\Delta y, z)\,.
    \label{eq:Idiv}
\end{align}
We emphasize that \cref{eq:DeltaSpoles} contains the divergence which was anticipated in our discussion of \cref{eq:TPforward,eq:TGforward}, and we can verify that the coefficient of this pole vanishes as $a\to0$ by noting that \cref{eq:cMpm} vanishes in this limit.
Having isolated the sole divergence in \cref{eq:DeltaSpoles}, we can use the forward or backward asymptotic measurement as a subtraction term to define the regularised integral,
\begin{align}
\begin{split}
    I^c_{\mathrm{reg.}}&=
    \int_{-\infty}^{\infty}
    \!
    \mathrm{d}\Delta y
    \int_0^{1}
    \!
    \mathrm{d}z
    \int_{0}^{\infty}
    \!
    \mathrm{d}y_t \, 
    e^{-4 \epsilon y_t}
    \left[
    \cM(y_t,\Delta y,z)-e^{4 y_t a \epsilon}\cM^{+}(\Delta y,z)
    \right]
    \cA^c(\Delta y, z)\,,\\
    &=
    \int_{-\infty}^{\infty}
    \!
    \mathrm{d}\Delta y
    \int_0^{1}
    \!
    \mathrm{d}z
    \int_{-\infty}^{0}
    \!
    \mathrm{d}y_t \, 
    e^{-4 \epsilon y_t}
    \left[
    \cM(y_t,\Delta y,z)-e^{4 y_t (2-a) \epsilon}\cM^{-}(\Delta y,z)
    \right]
    \cA^c(\Delta y, z)\,,
    \label{eq:Ireg}
\end{split}
\end{align}
which is finite in four dimensions. 
In terms of the integrals $I^c_{\mathrm{reg.}}$ and $I^c_{\mathrm{div.}}$, the soft function is given by 
\begin{align}
    &\Delta S_a^{\bare(2,c)}(\cTcut)=
    \left(\frac{\mu}{\cTcut} \right)^{4\epsilon}
    \bigg[\frac{1}{2(1-a)\epsilon} I^c_{\mathrm{div.}}+2 I^c_{\mathrm{reg.}}\bigg]\,,\\
    \begin{split}
    &
    \qquad=
    \frac{1}{2(1-a)\epsilon}
    I^{c,0}_{\mathrm{div.}}
    +2I^{c,0}_{\mathrm{reg.}}
    +\frac{1}{2(1-a)}\left( 
    4 I^{c,0}_{\mathrm{div.}}
    \log \left(\frac{\mu}{\cTcut}
    \right)
    +I^{c,1}_{\mathrm{div.}}
    \right)+\cO(\epsilon)\,.
    \label{eq:DeltaSbare_a}
\end{split}
\end{align}
In the second equality we have performed the expansions,
\begin{align}
    I^{c}_{\mathrm{div.}}=
    \sum_{n=0}^{\infty}
    \epsilon^n  I^{c,n}_{\mathrm{div.}}\,,
    \qquad \qquad
    I^{c}_{\mathrm{reg.}}=
    \sum_{n=0}^{\infty}
    \epsilon^n  I^{c,n}_{\mathrm{reg.}}\,.
\end{align}
In order to facilitate the analytic calculation of the integrals we perform a subsequent expansion in $a$,
\begin{align}
    I^{c}_{\mathrm{div.}}=
    \sum_{m=1}^{\infty}
    \sum_{n=0}^{\infty}
    \epsilon^n a^m I^{c,n,m}_{\mathrm{div.}}\,,
    \qquad \qquad
    I^{c}_{\mathrm{reg.}}=
    \sum_{m=0}^{\infty}
    \sum_{n=0}^{\infty}
    \epsilon^n a^m I^{c,n,m}_{\mathrm{reg.}}\,.
    \label{eq:I_en-am}
\end{align}
\begin{table}[htb]
\centering
$
\begin{array}{|c|c|c|}
\hline
I^{c,n,m}
& c=C_R n_f T_F &  c=C_R C_A
\\ \hhline{|=|=|=|}
\vphantom{\Big[}
\Idiv^{c,0,1}
& 
\frac{10}{3}
&     
-\frac{41}{3}+\frac{4}{3}\pi^2+8 \zeta_3
\\ \hline
\vphantom{\Big[}
\Idiv^{c,0,2}
& 
\frac{56}{45}
-
\frac{12}{5}\zeta_3
&   
-\frac{118}{45}
+\frac{1}{3}\pi^2
-\frac{24}{5}\zeta_3
\\ \hline
\vphantom{\Big[}
\Idiv^{c,0,3} 
&    
\frac{11}{54}
-
\frac{2}{9}\zeta_3
& 
-\frac{65}{108}
+\frac{1}{9}\pi^2
-\frac{8}{9}\zeta_3
+\frac{1}{90}\pi^4
\\ \hline
\vphantom{\Big[}
\Idiv^{c,0,4} 
&    
\frac{13}{210}
-
\frac{2}{15}\zeta_3
+
\frac{1}{7}
\zeta_5
& 
-\frac{19}{105}
+\frac{1}{24}\pi^2
-\frac{13}{30}\zeta_3
+\frac{1}{120}\pi^4
-\frac{4}{7}\zeta_5
\\ \hhline{|=|=|=|}
\vphantom{\Big[}
\Idiv^{c,1,1} 
&  
\frac{239}{9}
-\frac{32}{3}\zeta_3
& 
-\frac{1793}{18}
+\frac{16}{3}\pi^2
+\frac{160}{3}\zeta_3
+\frac{4}{9}\pi^4
\\ \hline
\vphantom{\Big[}
\Idiv^{c,1,2}
&  
\frac{2162}{225}
-\frac{2}{9}\pi^2
-\frac{212}{75}\zeta_3
-\frac{1}{9}\pi^4
& 
-\frac{2741}{225}
+\frac{4}{9}\pi^2
+\frac{2116}{75}\zeta_3
-\frac{26}{45}\pi^4
\\ \hline
\vphantom{\Big[}
\Idiv^{c,1,3} 
& 
\frac{467}{1620}
-\frac{1}{27}\pi^2
+\frac{202}{135}\zeta_3
-\frac{7}{405}\pi^4
&    
-\frac{4187}{3240}
+\frac{559}{135}\zeta_3
-\frac{31}{405}\pi^4
+\frac{2}{9}\pi^2\zeta_3
+\zeta_5
\\ \hhline{|=|=|=|}
\vphantom{\Big[}
\Ireg^{c,0,0}
&
-\frac{8}{3}
+\frac{16}{3}\zeta_3
& 
\frac{4}{3}
-\frac{44}{3}\zeta_3
+\frac{4}{15}\pi^4
\\ \hline
\vphantom{\Big[}
\Ireg^{c,0,1}
& 
\frac{25}{6}
-\frac{2}{3}\pi^2
-\frac{8}{3}\zeta_3
& 
-\frac{97}{12}
+\frac{1}{3}\pi^2
+\frac{34}{3}\zeta_3
-\frac{11}{45}\pi^4
\\ \hline
\vphantom{\Big[}
\Ireg^{c,0,2}
& 
\frac{76}{135}
-\frac{2}{9}\pi^2
+\frac{8}{225}\pi^4
&
-\frac{1021}{270}
+\frac{7}{36}\pi^2
+3\zeta_3
+\frac{13}{300}\pi^4  
\\ \hline
\vphantom{\Big[}
(*)\,\Ireg^{c,0,3}
&
\frac{161}{1296}
-\frac{1}{27}\pi^2
-\frac{11}{54}\zeta_3
+\frac{1}{162}\pi^4   
& 
-\frac{5885}{2592}
+\frac{5}{36}\pi^2
+\frac{127}{54}\zeta_3
-\frac{59}{3240}\pi^4  
-\frac{1}{9}\pi^2\zeta_3
+\frac{1}{3}\zeta_5
\\ \hline
\end{array}
$
\caption{The coefficients of $\epsilon^n a^m$ and colour structure $c$ in the expansion \cref{eq:I_en-am} of the integrals \cref{eq:Idiv} and \cref{eq:Ireg}. The coefficients marked with $(*)$ have been obtained using the PSLQ algorithm, as discussed in the main text.}
\label{tab:Icnm}
\end{table}
We have calculated the coefficients of $\Idiv^c$ up to $a^4$, and the coefficients of $\Ireg^c$ up to $a^3$.  The coefficients are are listed in \cref{tab:Icnm}. These coefficients were obtained by analytic solution of the integrals, with the exception of $\Ireg^{c,0,3}$. To achieve the analytic results, we utilized several results in ref.~\cite{hoffman2020logarithmicintegralszetavalues}. For $\Ireg^{c,0,3}$, we performed the integrals over $y_t$ and $\Delta y$ analytically. The remaining integral over $z$ was performed numerically to 100 significant figures. The first 50 significant figures were used to reconstruct the coefficient using the PSLQ algorithm~\cite{PSLQref}, with an ansatz that the integral evaluates to multiple zeta values (although several other transcendental numbers were also included in our basis to test this ansatz). The remaining 50 significant figures were used to verify the PSLQ result.

From \cref{eq:DeltaSbare_a} we determine that the renormalised differential soft function is
\begin{align}
    &\Delta S_a^{(2,c)}(\cT_S,\mu)=
    -\frac{2\Idiv^{c,0}}{(1-a)}\frac{1}{\mu}
    \cL_0\left(\frac{\cT_S}{\mu}\right)
    +\delta(\cT_S)
    \left(
    \frac{\Idiv^{c,1}}{2(1-a)}+2\Ireg^{c,0}\right)
    +\cO(\epsilon)\,,
    \label{eq:DeltaS_a_diff}
\end{align}
and the renormalised cumulant soft function is 
\begin{align}
    &\Delta S_a^{(2,c)}(\cTcut,\mu)=
    -\frac{2\Idiv^{c,0}}{(1-a)}\log \left(\frac{\cTcut}{\mu}
    \right)
    +
    \frac{\Idiv^{c,1}}{2(1-a)}+2\Ireg^{c,0}
    +\cO(\epsilon)\,.
    \label{eq:DeltaSren_a}
\end{align}

Writing the correction to the global (C-)angularity soft anomalous dimension as
\begin{equation}
\Delta \gamma_{a;S}^{(1,c)}(\cT_S,\mu)=\frac{4}{(1-a)}\Delta \Gamma_{\cusp}^{(1,c)}\frac{1}{\mu}\mathcal{L}_0\left(\frac{\cT_S}{\mu}\right)+\Delta \hat{\gamma}_{a;S}^{(1,c)}\delta(\cT_S)\,,
\label{eq:Delta_gamma}
\end{equation}
from \cref{eq:DeltaS_a_diff} we find
\begin{equation}
\Delta \Gamma_{\cusp}^{(1,c)}=0\,, \qquad
\Delta \hat{\gamma}_{a;S}^{(1,c)}=\frac{2}{(1-a)}\Idiv^{c,0}\,.
\label{eq:Delta_gamma_coeff}
\end{equation}
The vanishing of the cusp component is in line with our discussion of \cref{eq:gamma_global_nfTF} and \cref{eq:gamma_global_CA}.

To summarise, the full (C-)angularity anomalous dimension is given by
\begin{equation}
    \gamma_{a;S}^{(1)}(\cT_S, \mu)=C_R \sum_{c \in \{C_A, n_f T_F\}}
    c
    \left(
    \gamma_{\Sigma,a;S}^{(1, c)}(\cT_S, \mu)
    +\Delta \hat{\gamma}_{a;S}^{(1,c)} \delta(\cT_S)
    \right)\,,
    \label{eq:gamma1a}
\end{equation}
with constituents given in \cref{eq:gamma_global_CA,eq:gamma_global_nfTF,eq:Delta_gamma_coeff}. Comparing to the same quantity computed in ref.~\cite{Bauer:2020npd}, we report a disagreement with their $C_R C_A$ non-cusp coefficient at order $a^2$. This disagreement is numerically small (0.2 per mille in the coefficient) and phenomenologically irrelevant.

\begin{figure}
\begin{subfigure}{.49\textwidth}
  \centering
  \includegraphics[width=1.0\linewidth]{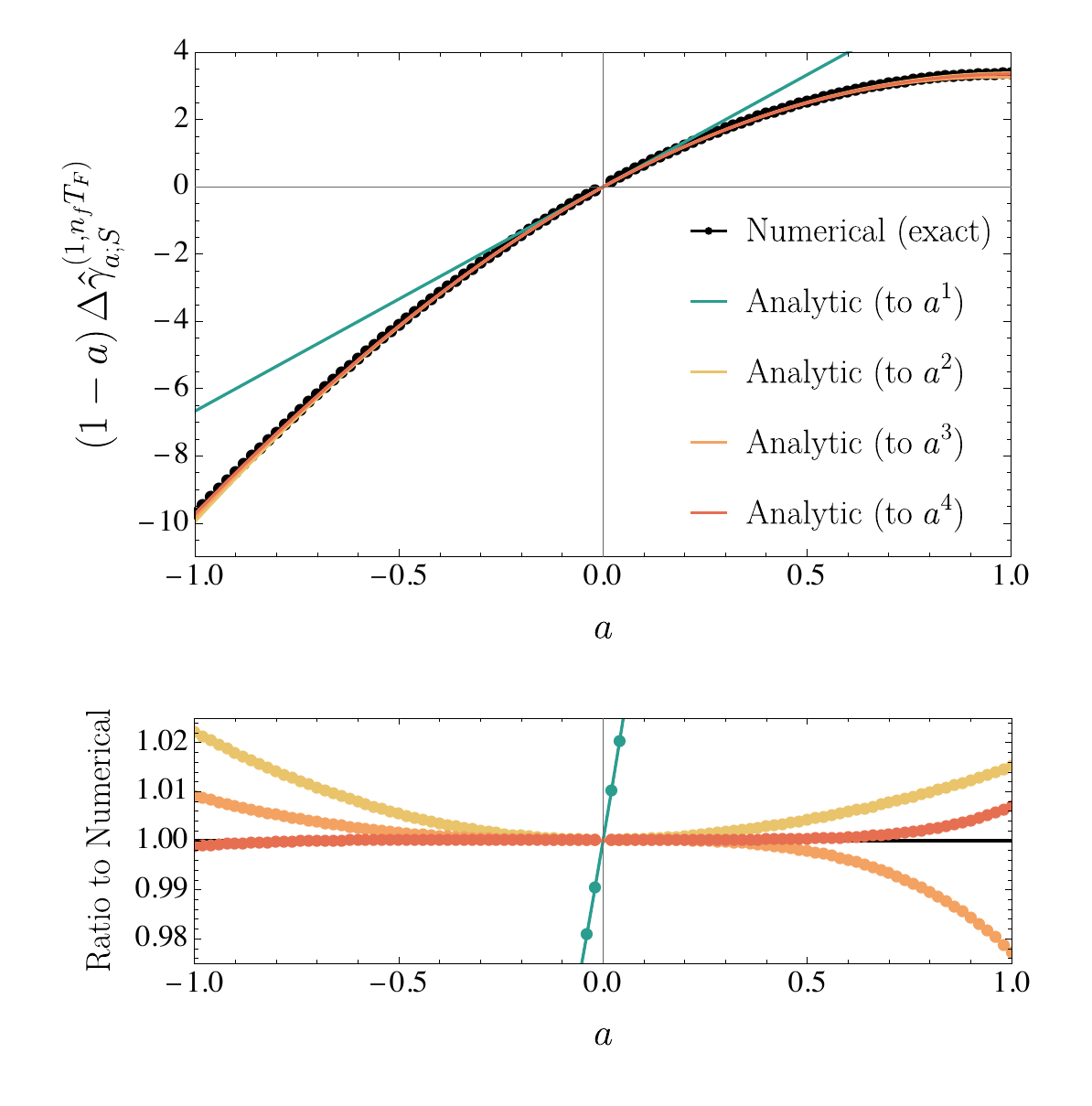}  
  \caption{$c= n_f T_F$}
  \label{fig:Dg_nfTF}
\end{subfigure}
\begin{subfigure}{.49\textwidth}
  \centering
  \includegraphics[width=1.0\linewidth]{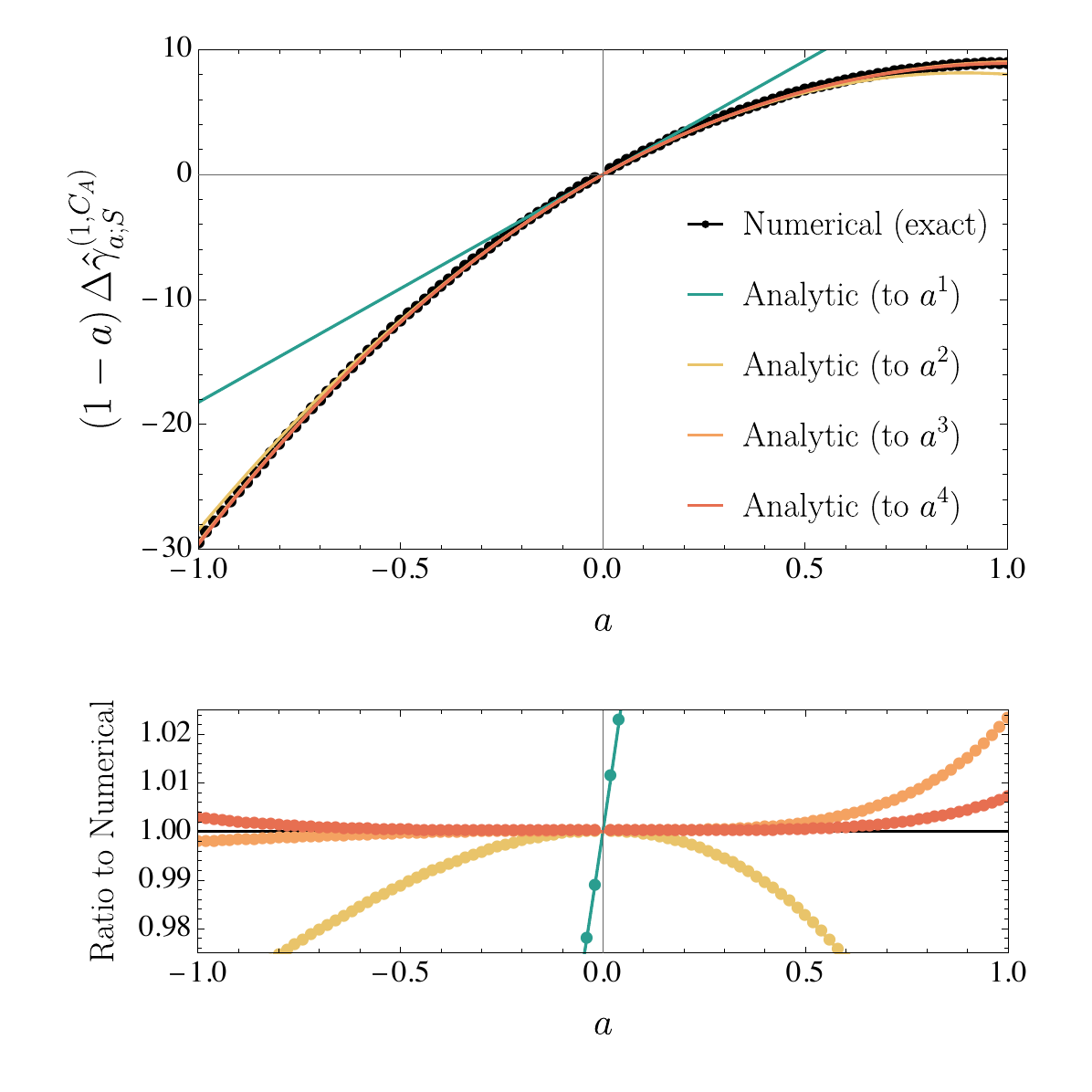}
  \caption{$c= C_A$}\label{fig:Dg_CA}
\end{subfigure}
\caption{A comparison of the exact $a$-dependence of $\Delta \hat{\gamma}_{a;S}^{(1,c)}$, computed numerically, with a Laurent expansion in $a$ to different orders in $a$, the coefficients of which were computed analytically and are tabulated in \cref{tab:Icnm}.}
\label{fig:Dg}
\end{figure}

To assess the accuracy of our expansion of $\Delta \hat{\gamma}_{a;S}^{(1,c)}$ in $a$, in \cref{fig:Dg} we have compared this analytic expansion to a numerical calculation of the same quantity. We see that the expansion to $\cO(a^4)$ remains within a few permille agreement with the exact result across the range $a \in (-1,1)$. Beyond this, the $\cO(a^4)$ expansion remains within 2 percent agreement in the range $a \in (-1.8,-1)$ for both colour channels.

\begin{figure}
\begin{subfigure}{.49\textwidth}
  \centering
  \includegraphics[width=\linewidth]{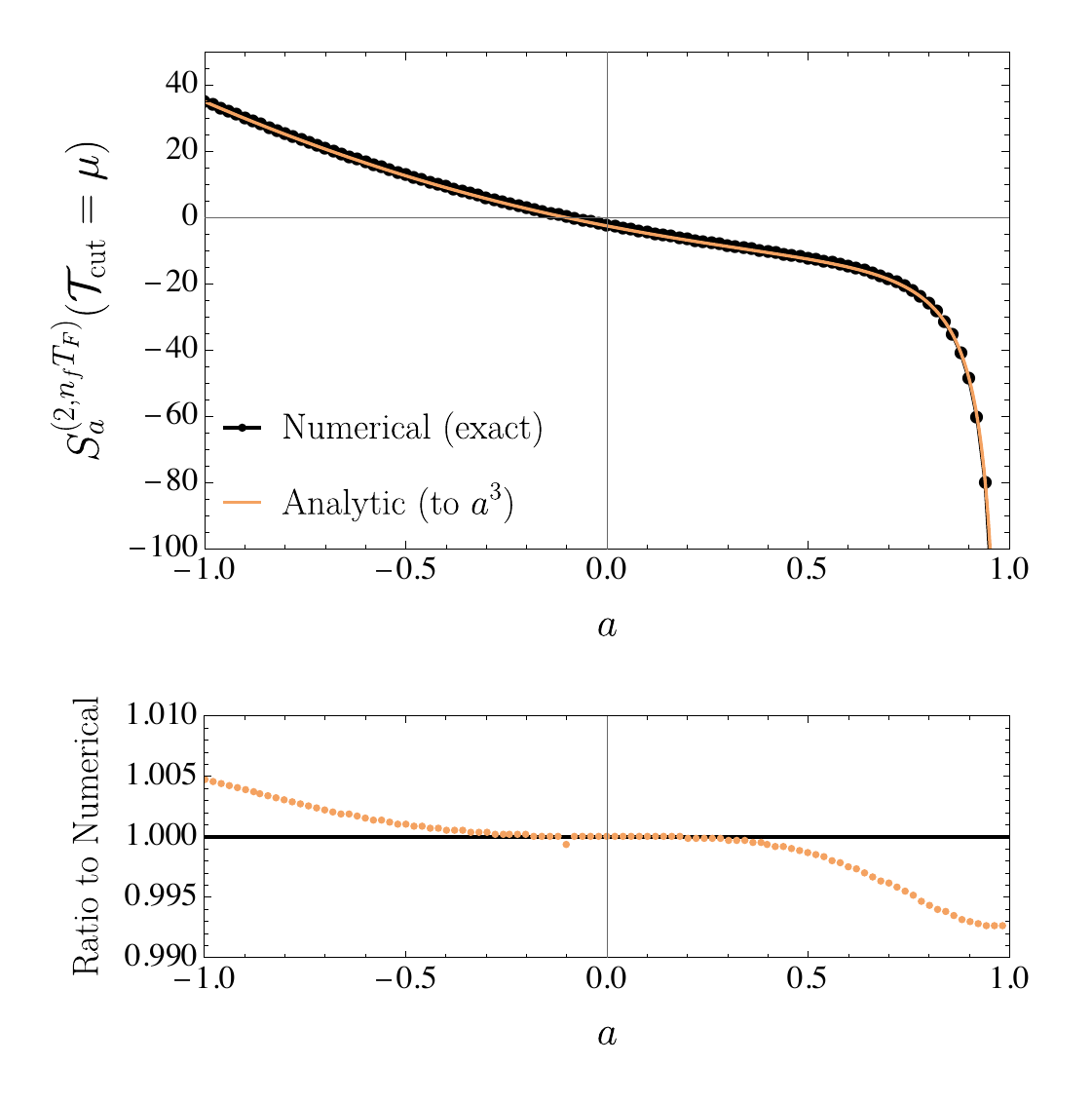}  
  \caption{$c= n_f T_F$}
  \label{fig:S2a_nfTF}
\end{subfigure}
\begin{subfigure}{.49\textwidth}
  \centering
  \includegraphics[width=\linewidth]{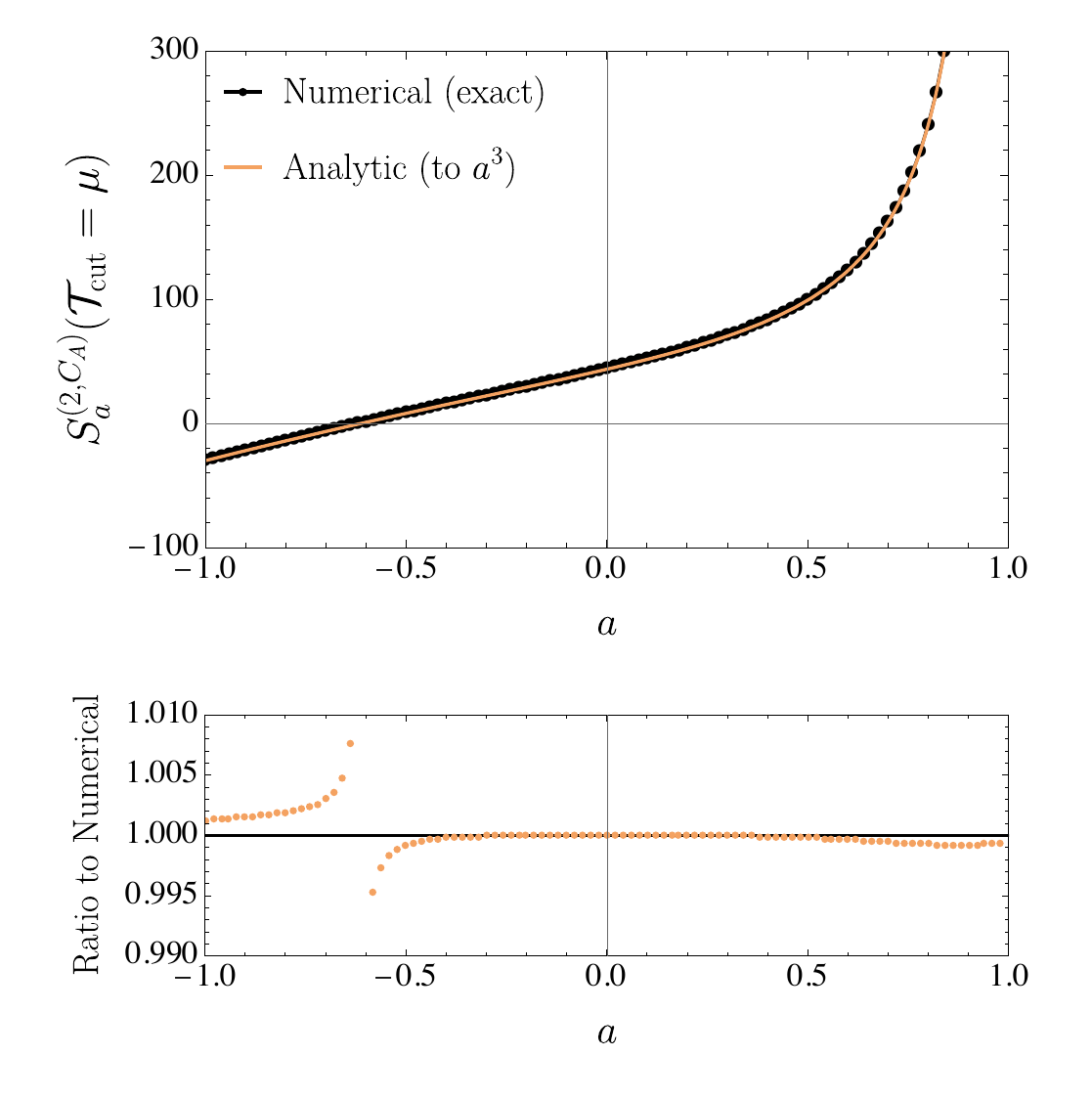}
  \caption{$c= C_A$}\label{fig:S2a_CA}
\end{subfigure}
\caption{A comparison of the exact $a$-dependence of $S^{(2,c)}_{a}$, where the $\Delta S^{(2,c)}_{a}$ component has been computed numerically, to a fully analytic calculation, where $\Delta S^{(2,c)}_{a}$ has been computed as a Laurent expansion in $a$ to $\cO(a^3)$. Here we have set $\mu=\cTcut$ so only the non-logarithmic terms contribute.}
\label{fig:S2a}
\end{figure}

To assess the phenomenological impact of the $a$-expansion of the non-logarithmic terms in \cref{eq:DeltaSren_a}, we assemble the full soft function for $c \in \{C_A, n_f T_F\}$ according to 
 \begin{align}
S_a^{(2,c)}= 
S_{\Sigma,a}^{(2,c)} + \Delta S_a^{(2,c)}\,.
\label{eq:S_split_ren}
\end{align}   
In \cref{fig:S2a} we compare the expansion of the full cumulant soft function up to $\cO(a^3)$ with an exact numerical calculation. We choose $\mu=\cTcut$ such that the comparison in \cref{fig:S2a} is independent of the quantities compared in \cref{fig:Dg}. As above, we find the expansion to $\cO(a^3)$ is accurate across the plotted range of $a$. In particular, the expansion is accurate to within 2 permille in the range $a \in (-0.5,0.5)$ for both colour channels. Furthermore, we report that the $\cO(a^3)$ expansion remains within 2 percent agreement in the range $a \in (-2.2,1)$ for both colour channels. We conclude that the coefficients collected in \cref{tab:Icnm} are sufficient for phenomenological purposes across a wide range of $a$, although if necessary, the expansion could be pushed to several higher orders in $a$ by utilising the PSLQ algorithm in the manner described above.

\subsection{Assembling the C-angularity soft function}

In terms of the various pieces computed in this section, the complete C-angularity soft function at N$^2$LO is given by
\begin{equation}
    S_a^{(2)}(\cTcut,\mu)=C_R
    \left[
    C_R \, S_a^{(2,C_R)}(\cTcut,\mu)
    + 
    \hspace{-2 em}
    \sum_{c \in \{C_A, n_f T_F\}}
    \hspace{-1 em}
    c
    \left(
    S_{\Sigma,a}^{(2, c)}(\cT_S, \mu)
    +\Delta S_{a}^{(2,c)}(\cT_S, \mu)
    \right)
    \right]\,.
    \label{eq:S2a}
\end{equation}
Expressions for the constituent terms are given in \cref{eq:Sa2CRCR,eq:S_global_nfTF,eq:S_global_CA,eq:DeltaS_a_diff}. A \texttt{Mathematica} \cite{Mathematica} implementation of the soft function is provided in the ancillary files accompanying this paper. Two different implementations of the integrals $\Idiv$ and $\Ireg$ are provided: a fully analytic implementation of the expansion around $a=0$ utilising the coefficients of \cref{tab:Icnm}, as well as a numerical implementation of the integrals for use at arbitrary $a\neq 1$.

Several checks have been performed on \cref{eq:S2a}. As discussed in \cref{sec:global}, \cref{eq:S2a} agrees with the N$^2$LO soft threshold function for $a=2$~\cite{Belitsky:1998tc,Becher:2007ty}. As discussed in \cref{sec:Delta}, the C-angularity soft anomalous dimension agrees with the angularity soft anomalous dimension of ref.~\cite{Bauer:2020npd} up to a minor discrepancy at order $a^2$. 
Finally, we confirm that upon taking a Laplace transform and setting $a=0$, \cref{eq:S2a} reproduces the N$^2$LO C-parameter soft function which was previously computed analytically in ref.~\cite{Bell:2018oqa}.

\section{Conclusion}
\label{sec:conclusion}

In this paper, we have calculated the NLO and NNLO soft function for a family of event shapes we call C-angularity, which depend on a real parameter $a<2$. Apart from a brief discussion in ref.~\cite{Lee:2006nr}, C-angularity has not previously been studied in the literature. 
For the special cases of $a=0$, 1 and 2, the C-angularity measurement reduces to C-parameter, broadening and  threshold respectively.
Furthermore, C-angularity coincides with conventional angularity in the collinear limit, so their anomalous dimensions also coincide. The results presented here, when combined with existing results in the literature \cite{Hornig:2009vb,Bell:2018gce, Bell:2021dpb, Zhu:2021xjn,Bell:2024lwy}, enable the resummation of C-angularities in $e^+e^-$ collisions and DIS up to the NNLL$^\prime$ order, and are a required ingredient for the N$^3$LL resummation. Comparing these precise resummed results to data for a variety of values of $a$ should be useful in terms of extracting the strong coupling constant $\alpha_s$ and further exploring/determining the impact of nonperturbative power corrections on event shapes.

In terms of the computation of the soft function, the advantage of the C-angularity measurement over the conventional angularity one is that it is a continuously differentiable function of the momenta, with no abrupt transitions inside the phase space. Such transitions can be complicated to handle analytically \cite{Baranowski:2021gxe}, and can potentially cause a numerical calculation to be less stable \cite{Bell:2018oqa}. For the NNLO calculation of the nontrivial `correlated' colour channels, we utilized a strategy of dividing the calculation into two pieces -- one term in which the C-angularity measurement is applied to the total momentum of emitted particles, and then a correction term to the full C-angularity (following the same general strategy as refs.~\cite{Bauer:2020npd,Jouttenus:2011wh,Banfi:2014sua,Gangal:2016kuo,Abreu:2022sdc, Abreu:2022zgo, Abreu:2026FutureJV, Buonocore:2026FutureTau}). The former piece may be straightforwardly extracted from results in the literature, whilst the latter is easier to compute than the full soft function owing to the fact that it can only contain a single divergence. 

We obtained analytical results for the NNLO C-angularity soft function as an expansion in $a$ around $a=0$. For the (C-)angularity anomalous dimension, we have been able to improve on the known accuracy of the angularity anomalous dimension \cite{Bauer:2020npd} by two orders in an expansion around $a=0$, up to $\cO(a^4)$. We have computed the non-logarithmic terms in the NNLO soft function to $\cO(a^3)$, which is within 2 permille accuracy in the range $a \in (-0.5,0.5)$ and within 2 percent  accuracy in the range $a \in (-2.2,1)$. These results should be sufficient for phenomenological purposes, but if necessary it is readily possible to obtain higher order terms in the $a$-expansion (at least numerically). We also provide a numerical implementation of \cref{eq:Idiv} and \cref{eq:Ireg} for use at arbitrary $a\neq 1$.

\section*{Acknowledgments}

The work of EB and JRG has been supported by the Royal Society through Grant 
URF\textbackslash{}R1\textbackslash{} 201500. JRG would like to thank Pier Monni for useful discussions. The Feynman diagrams in this paper were produced with the aid of {\tt FeynCraft} \cite{Gaunt:2025peq} and the {\tt TikZ-Feynman} package \cite{Ellis:2016jkw}.

\appendix
\section{Kinematic conventions}
\label{sec:kin}
We decompose an arbitrary four-vector $k^\mu$ as
\begin{equation}
    k^\mu = \frac{1}{2}\left( k^+ n^\mu +k^- \bar{n}^\mu\right)+k_\perp^\mu\,,
\end{equation}
where $n^\mu$ and $\bar{n}^\mu$ are light-like reference vectors along the beam directions which satisfy
\begin{align}
    n^2=\bar{n}^2=0\,, \qquad n \cdot \bar{n}=2\,.
\end{align}
We take these reference vectors to be along the beam directions,
\begin{align}
    n=(2,0,0,2)\,, \qquad  \bar{n}=(2,0,0,-2)
\end{align}
such that
\begin{align}
    k^\pm=k^0 \pm k^z\,.
\end{align}
We denote by $\vec{k}_\perp$ the Euclidean two-vector comprising the transverse components of the Lorentzian four-momentum $k_\perp$, and we denote by $k_T$ the scalar magnitude of $k_\perp$. These objects are related via
\begin{equation}
   -k_\perp \cdot k_\perp = \vec{k}_\perp \cdot \vec{k}_\perp 
   =|\vec{k}_\perp|^2 =k_T^2 \ge 0\,,
\end{equation}
where we use the symbol ``$\cdot$" to denote either the Lorentzian or Euclidean inner product depending on the context. 

\section{Perturbative expansions}
\label{sec:expansion}
We adopt standard dimensional reduction with $4-2\epsilon$ dimensions, and we work in the $\msbar$ scheme. We express all quantities in terms of the renormalised coupling, for which we need the one-loop relation to the bare coupling,
\begin{equation}
   \alpha_s^\bare =\alpha_s(\mu) \left(\frac{\mu^2 e^{\gamma_E}}{4\pi}\right)^\epsilon\left[
   1-\frac{\beta_0}{\epsilon}\frac{\alpha_s(\mu)}{4\pi}+\cO(\alpha_s(\mu)^2)
   \right]\,.
\end{equation}
The renormalised coupling satisfies
\begin{equation}
   \mu \frac{\mathrm{d}}{\mathrm{d} \mu} \alpha_s(\mu)=-2 \epsilon\alpha_s(\mu)
   -2 \beta_0 \frac{\alpha_s(\mu)^2}{4 \pi} + \cO(\alpha_s(\mu)^3)\,,
\end{equation}
with
\begin{equation}
   \beta_0 =\frac{11 C_A-4 n_f T_F}{3}\,.
\end{equation}
We expand the cumulant and differential soft functions and renormalisation functions in the renormalised coupling:
\begin{align}
    S_a^{\bare}=
    \sum_{n=0}^{\infty}
    \left(
\frac{\alpha_s(\mu)}{4 \pi}
\right)^{n}
S_a^{\bare(n)}
\,,\qquad
    S_a=
    \sum_{n=0}^{\infty}
    \left(
\frac{\alpha_s(\mu)}{4 \pi}
\right)^{n}
S_a^{(n)}\,,
\\
Z_a^{\bare}=
    \sum_{n=0}^{\infty}
    \left(
\frac{\alpha_s(\mu)}{4 \pi}
\right)^{n}
Z_a^{\bare(n)}
\,,\qquad
Z_a=
    \sum_{n=0}^{\infty}
    \left(
\frac{\alpha_s(\mu)}{4 \pi}
\right)^{n}
Z_a^{(n)}\,.
\end{align}
As is conventional, we start the expansion of anomalous dimensions at one higher power of the coupling,

\begin{align}
\gamma_{a;S} &= 
\sum_{n=0}^{\infty}
    \left(
\frac{\alpha_s(\mu)}{4 \pi}
\right)^{n+1}
\gamma_{a;S}^{(n)}\,,
\\
\hat{\gamma}_{a;S} [\alpha_s(\mu)]&= 
\sum_{n=0}^{\infty}
    \left(
\frac{\alpha_s(\mu)}{4 \pi}
\right)^{n+1}
\hat{\gamma}_{a;S}^{(n)}\,,
\\
\Gamma_\cusp [\alpha_s(\mu)]&=
\sum_{n=0}^{\infty}
\left(
\frac{\alpha_s(\mu)}{4 \pi}
\right)^{n+1}
\Gamma_{\cusp}^{(n)}\,.
\end{align}
To check our results, we need the first two cusp anomalous dimension coefficients~\cite{Korchemsky:1987wg}
\begin{align}
\Gamma_{0}&= 4 \label{eq:Gamma0}\,,\\
\Gamma_{1}&= 4 \left[
C_A\left(\frac{67}{9}-\frac{\pi^2}{3}\right)
-\frac{20}{9}T_F n_f
\right]\,,\label{eq:Gamma1}
\end{align}
where $\Gamma_n=\Gamma_{\cusp}^{(n)}/C_R$ are the standard representation-independent coefficients.

\section{Plus distributions}

We use the standard distributions
\begin{equation}
\cL_n(x)=\left[\frac{\theta(x) \ln^n x}{x}\right]_{+}=\lim _{\epsilon \rightarrow 0} \frac{\mathrm{d}}{\mathrm{d} x}\left[\theta(x-\epsilon) \frac{\ln^{n+1} x}{n+1}\right]\,.
\end{equation}
We need the distributional identity
\begin{equation}
\frac{\theta(x)}{x^{1-\epsilon}}=\frac{1}{\epsilon} \delta(x)+\sum_{n=0}^{\infty} \frac{\epsilon^n}{n!} \cL_n(x)\,,
\label{eq:dist}
\end{equation}
the derivatives
\begin{align}
    \mu \frac{\mathrm{d}}{\mathrm{d} \mu} \frac{1}{\mu} \cL_n\left(\frac{z}{\mu} \right)&=
    -n\frac{1}{\mu}\cL_{n-1}\left(\frac{z}{\mu} \right) \qquad \forall n>0\,,
    \\
    \mu \frac{\mathrm{d}}{\mathrm{d} \mu} \frac{1}{\mu} \cL_0\left(\frac{z}{\mu} \right)&=-\delta(z)\,,
\end{align}
and the integrals
\begin{align}
\int_0^{z_\cut} \frac{\mathrm{d} z}{\mu} \cL_{n}\left(\frac{z}{\mu} \right)&=\frac{1}{n+1}\log\left(\frac{z_\cut}{\mu} \right)^{n+1} \qquad \forall n\ge0\,.
\end{align}

\bibliography{refs.bib}

\end{document}